\documentclass[aps%
,pre%
,twocolumn%
,showpacs%
,floats
,amssymb%
]{revtex4}
\usepackage{amsmath}%
\usepackage{graphicx}%
\usepackage{color}%

\newcommand{\comment}[1]{}

\begin{document}
\title{Geometry and Dynamics for Hierarchical Regular Networks }
\author{Stefan Boettcher}
\homepage{http://www.physics.emory.edu/faculty/boettcher/}
\author{Bruno Goncalves}
\homepage{http://www.bgoncalves.com/index.shtml}
\author{Julian Azaret}
\affiliation{Emory University, Dept. of Physics, Atlanta, GA 30322}
\date{\today}

\begin{abstract}
The recently introduced hierarchical regular networks HN3 and HN4 are
analyzed in detail. We use renormalization group arguments to show
that HN3, a 3-regular planar graph, has a diameter growing as
$\sqrt{N}$ with the system size, and random walks on HN3 exhibit
super-diffusion with an anomalous exponent
$d_{w}=2-\log_{2}(\phi)\approx1.306$, where
$\phi=\left(\sqrt{5}+1\right)/2=1.618\ldots$ is the {}``golden
ratio.'' In contrast, HN4, a non-planar 4-regular graph, has a
diameter that grows slower than any power of $N$, yet, fast than any
power of $\ln N$. In an annealed approximation we can show that
diffusive transport on HN4 occurs ballistically ($d_{w}=1$). Walkers
on both graphs possess a first-return probability with a power law
tail characterized by an exponent $\mu=2-1/d_{w}$. It is shown
explicitly that recurrence properties on HN3 depend on the starting
site.
\end{abstract}
\maketitle

\section{Introduction}

Networks with a sufficiently intricate structure to exhibit not-trivial
properties for statistical models but sufficiently simple to reveal
analytical insights are few and far between. Familiar examples are
random graphs \cite{ER,Bollobas}, the hierarchical lattices \cite{Berker79}
originating in the Migdal-Kadanoff bond-moving scheme \cite{Migdal76,Kadanoff76},
fractal lattices \cite{Mandelbrot82}, or scale-free networks \cite{Barabasi01,Andrade05,Zhang07}.
Scale-free networks and random graphs can elucidate mean-field properties
only, whereas hierarchical lattices often provide excellent results
for complex statistical models in low dimensions \cite{Southern77}
but do not possess a mean-field limit. Fractal lattices provide tantalizing
access to dynamical systems in non-integer dimensions but can not
be tuned across dimensions. From the perspective of statistical physics,
it could be desirable to have a network that combines solvability
for low-dimensional systems with mean-field properties in such a way
that one could interpolate between either extreme. 

We have recently introduced a set of networks that overlay a lattice
backbone with regular long-range links \cite{SWPRL,SWN}, similar
to small-world networks \cite{Watts98} but hierarchical and without
randomness. These networks have a recursive construction but retain
a fixed, regular degree. The hierarchical sequence of long-distance
links occur in a pattern reminiscent of the tower-of-hanoi sequence
\cite{SWN}. Therefore, we have dubbed them {}``Hanoi-Networks''
and abbreviated them as HN3 and HN4, since one is 3-regular and the
other 4-regular. While almost identical, both types of networks lead
to very distinct behaviors, as revealed by our studies here. Future
work will focus on detailed studies of Ising models on these networks.
In Ref.~\cite{SWPRL} we have shown already that spin models on HN4
have the desired properties mentioned above. Here, we analyze in great
detail diffusive transport on these networks, for which especially
HN3 proves to possess a rich behavior. 

The paper is structured as follows. In the next section, we introduce
the networks and discuss various key design aspects. In Sec.~\ref{sec:Graph-Structure},
we explore geometric aspects of the networks such as their diameter.
Sec.~\ref{sec:Random-Walks} contains our analysis of random walks
on both networks, starting with a review of the dynamic renormalization
group in the context of a simple one-dimensional random walk which
facilitates an efficient discussion of the equivalent but far more
involved calculation for HN3. Unfortunately, the same approach does
not seem to apply to HN4, so we conclude the section with a derivation
of a moment-generating equation in Fourier space for walks on HN4
directly from the master equation and an annealed approximation. We
conclude this paper in Sec.~\ref{sec:Conclusions}, indicating various
future projects that can be developed from the work presented here.

\section{Network Design\label{sec:Network-Design}}

Both networks we are discussing in this paper consist of a simple
geometric backbone, a one-dimensional line of $N=2^{k}$ sites, either
infinite $\left(-2^{k}\leq n\leq2^{k},\, k\to\infty\right)$, semi-infinite
$\left(0\leq n\leq2^{k},\, k\to\infty\right)$, or closed into a ring.
Each site on the one-dimensional lattice backbone is connected to
its nearest neighbor. (In general, the following procedure also works
with a higher-dimensional lattice.) 

\begin{figure}
\includegraphics[clip,scale=0.3]{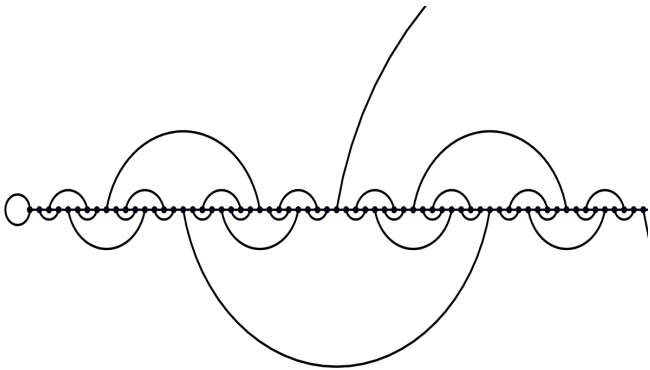} 
\caption{Depiction of HN3 on a semi-infinite line. The leftmost
site here is $n=0$, which requires special treatment. The entire
network can be made 3-regular with a self-loop at $n=0$. Note that
HN3 is planar.}
\label{fig:3hanoi}
\end{figure}

\begin{figure}
\includegraphics[clip,scale=0.35,angle=90]{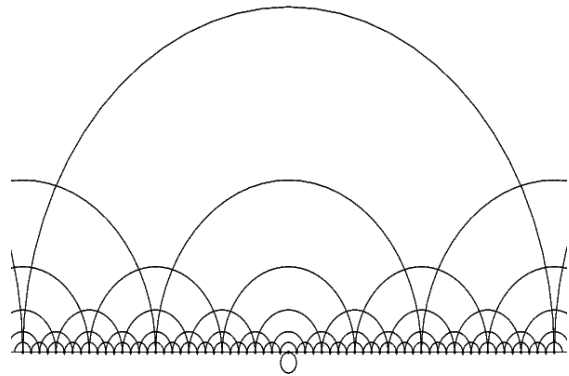}
\caption{Depiction of HN4 on an infinite line. The center
site is $n=0$, which requires special treatment. The entire network
becomes 4-regular with a self-loop at $n=0$. Note that HN4 is non-planar.}
\label{fig:4hanoi}
\end{figure}

To generate the small-world hierarchy in these graphs, consider parameterizing
any integer $n$ (except for zero) \emph{uniquely} in terms of two
other integers $(i,j)$, $i\geq0$, via
\begin{eqnarray}
n & = & 2^{i}\left(2j+1\right).
\label{eq:numbering}
\end{eqnarray}
Here, $i$ denotes the level in the hierarchy whereas $j$ labels
consecutive sites within each hierarchy. For instance, $i=0$ refers
to all odd integers, $i=1$ to all integers once divisible by 2 (i.
e., 2, 6, 10,...), and so on. In these networks, aside from the backbone,
each site is also connected with (one or both) of its nearest neighbors
\emph{within} the hierarchy. For example, we obtain the 3-regular
network HN3 (best done on a semi-infinite line) by connecting first
all nearest neighbors along the backbone, but in addition also 1 to
3, 5 to 7, 9 to 11, etc, for $i=0$, next 2 to 6, 10 to 14, etc, for
$i=1$, and 4 to 12, 20 to 28, etc, for $i=2$, and so on, as depicted
in Fig.~\ref{fig:3hanoi}. Correspondingly, HN4 is obtained in the
same manner, but connecting to \emph{both} neighbors in the hierarchy,
i. e., 1 to 3, 3 to 5, 5 to 7, etc, for $i=0$, 2 to 6, 6 to 10, etc,
for $i=1$, and so forth. For this network it is clearly preferable
to extend the line to $-\infty<n<\infty$ and also connect -1 to 1
, -2 to 2, etc, as well as all negative integers in the above pattern,
see Fig.~\ref{fig:4hanoi}. The site with index zero, not being covered
by Eq.~(\ref{eq:numbering}), is clearly a special place, either on
the boundary of the HN3 or in the center of the HN4. It is easy to
generalize these graphs, for instance, by putting the structure on
a ring with periodic boundary conditions, which may require special
treatment of the highest level in the hierarchy on finite rings.


\begin{figure}
\includegraphics[clip,scale=0.35,angle=-90]{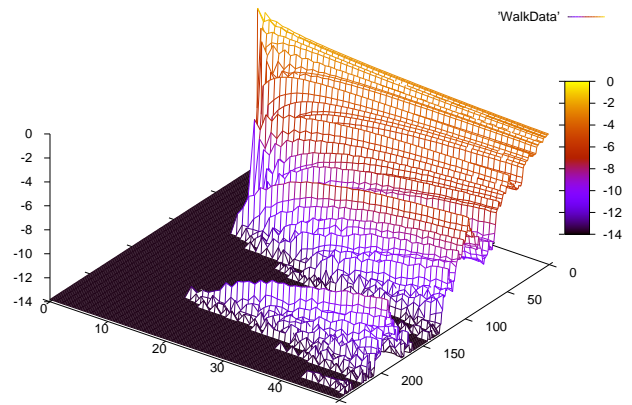} 
%
\caption{Plot of the probability $P_{l,t}$ of a random
walker to reside at a site $l$ at time $t$ after starting at the
origin $l=0$ obtained by numerical simulations on HN3. The scale
of the vertical axis refers to $\ln P_{l,t}$ to increase the visibility.
Note the highly non-homogeneous evolution, which is tied to fractal
distance-dependence also apparent in Fig.~\ref{fig:3dia} due to the
hierarchy of long-range links. Those links allow for walkers to appear
suddenly at certain sites such as $l=3\times2^{i}=6,12,24,48,96,192,\ldots$
long before ordinary diffusion along the backbone itself would provide
for. Those onset-points for the most rapidly progressing walker should
follow an envelope function that is linear ({}``ballistic'') in
a space-time plot, as walkers diffuse ordinarily with hopping-distance,
$d\sim\sqrt{t}$, but by Eq.~(\ref{eq:3dia}), $d\sim\sqrt{l}$, hence,
$l\sim t$. Average walkers progress slower than ballistic, though,
but still faster than diffusion.}
\label{fig:HN3RW-3D}
\end{figure}

Random walks on these networks have fascinating properties due to
their fractal nature and their long-range links as shown, for instance,
in Fig.~\ref{fig:HN3RW-3D}. %

\section{Network Geometry\label{sec:Graph-Structure}}

\subsection{Distance Measure on HN3\label{sub:Distance-measure-on}}

\begin{figure}
\includegraphics[clip,scale=0.35,angle=90]{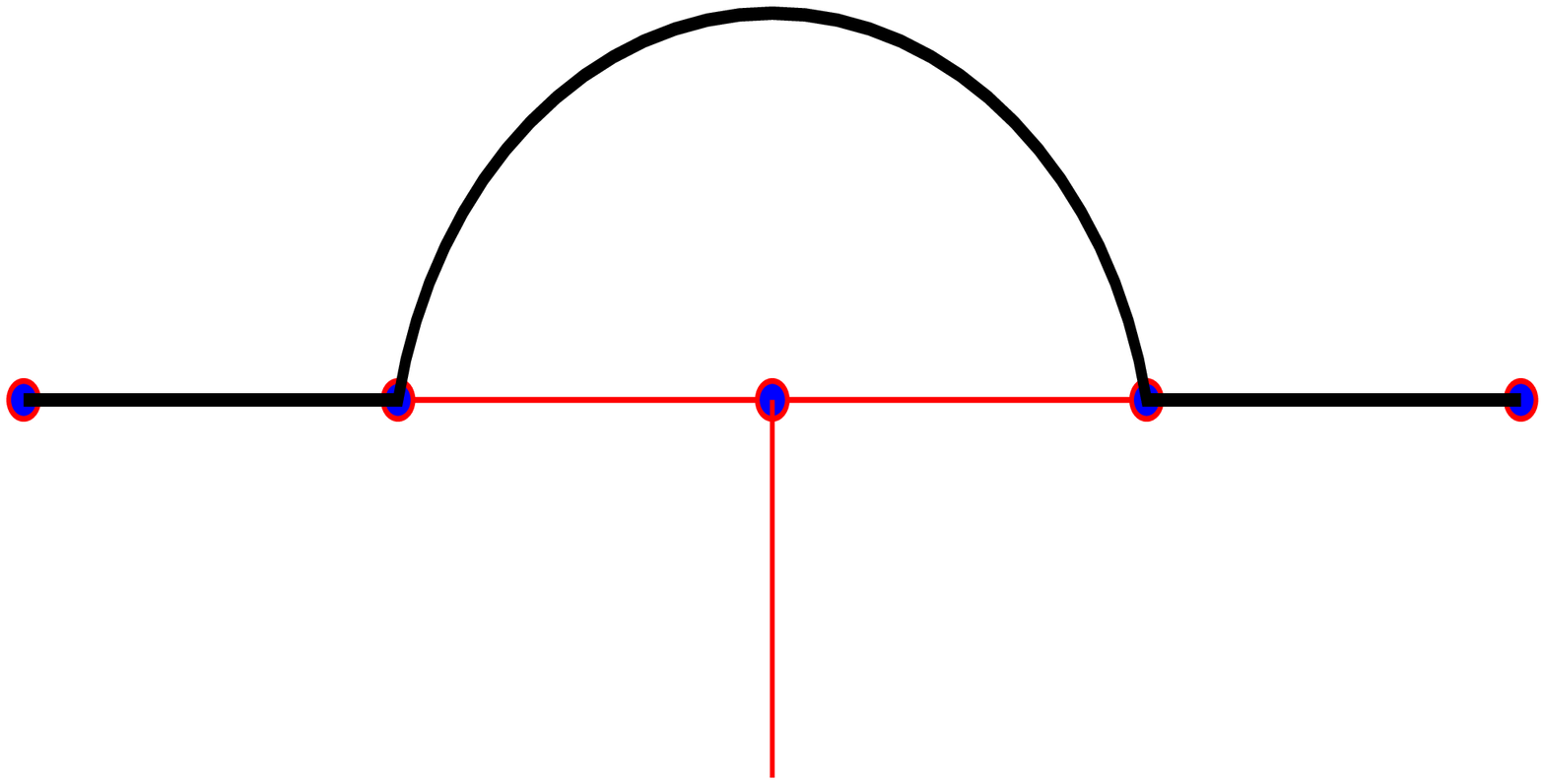} 
\includegraphics[clip,scale=0.35,angle=90]{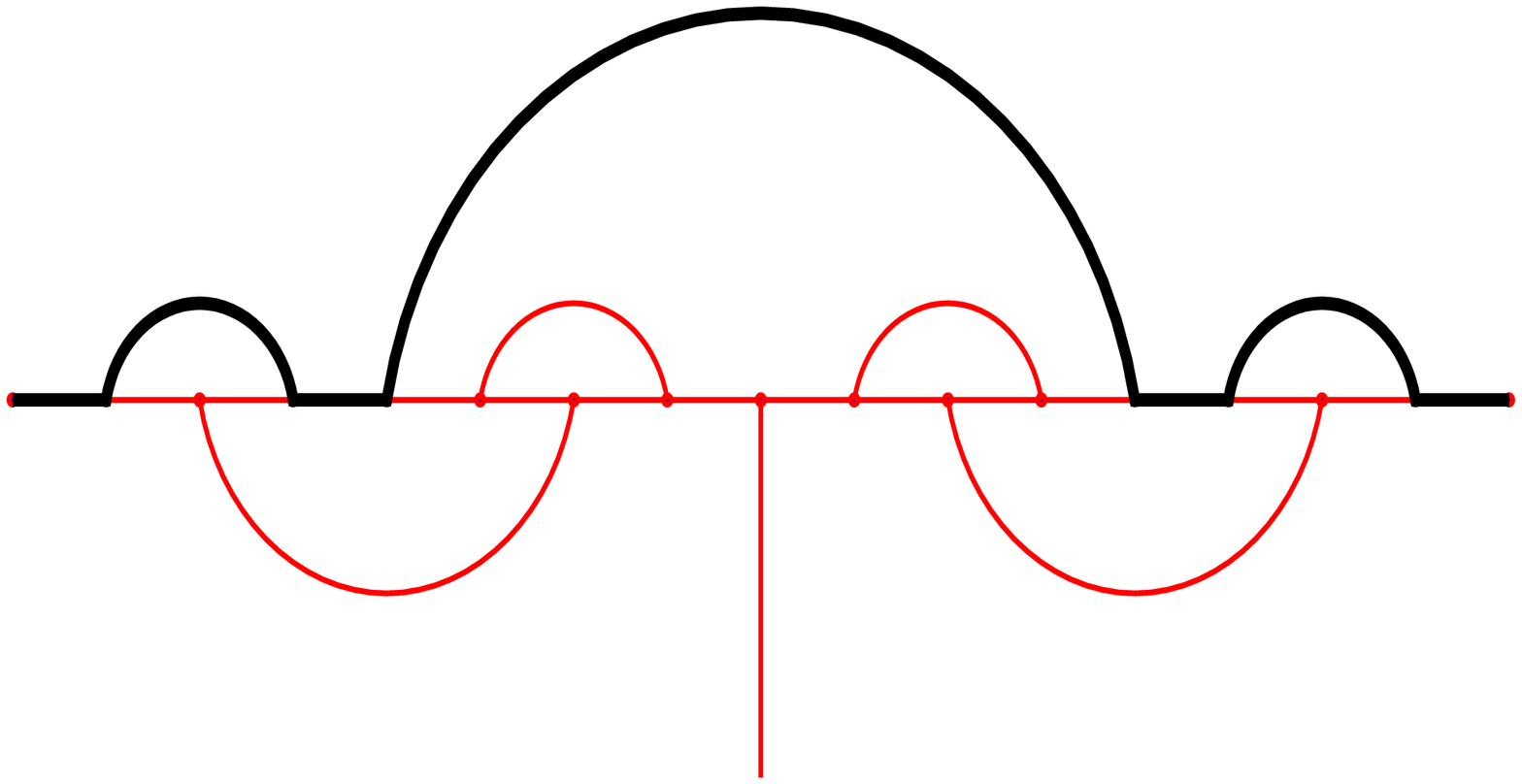} 
\includegraphics[clip,scale=0.35,angle=90]{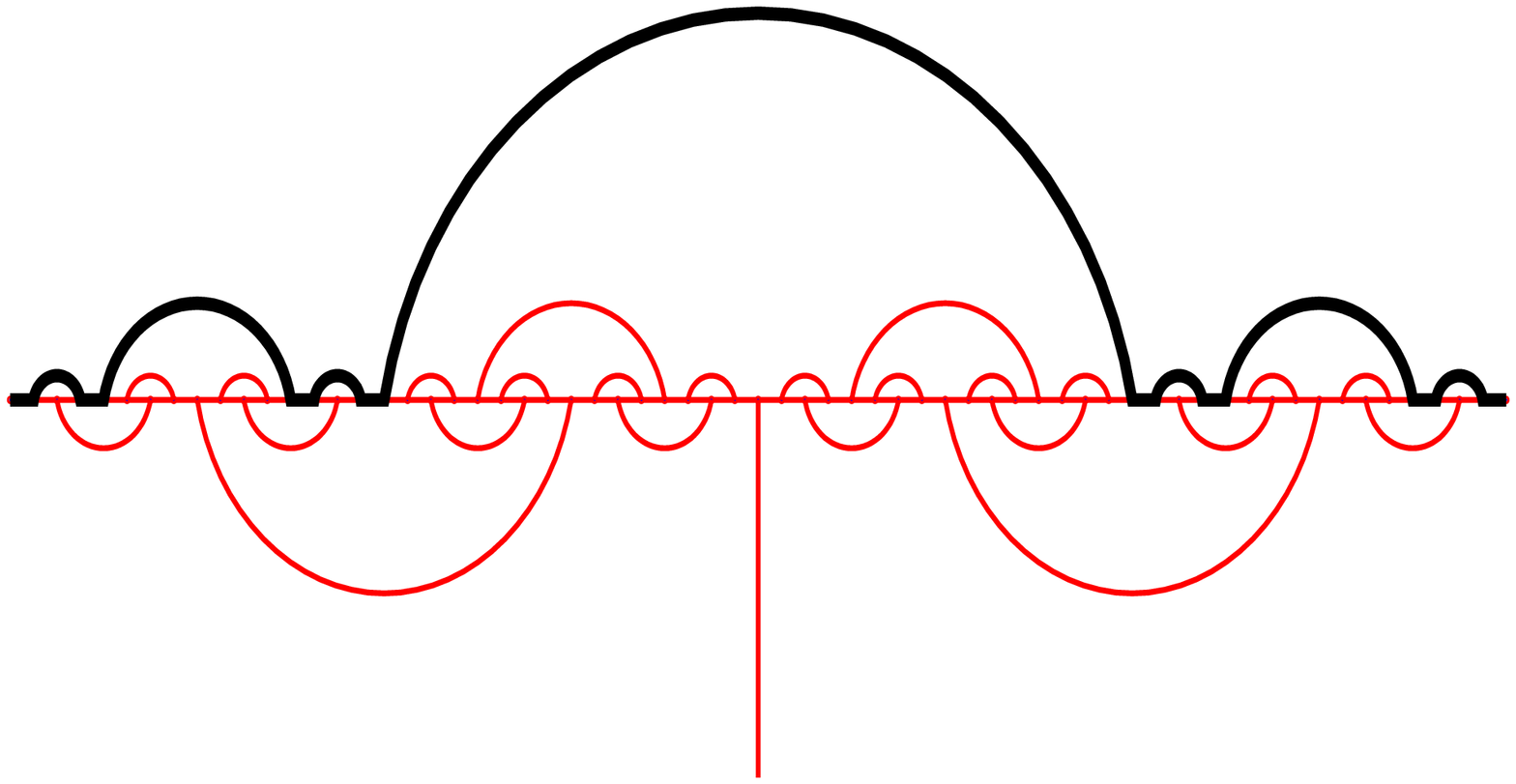} 
\caption{
Sequence of shortest end-to-end paths (=diameter,
thick black lines) for HN3 of size $N=2^{k}$, $k=2,4,8$. (Note that
every second step in the hierarchical development has been omitted.
At level $k=3,5,\ldots$ the shortest path is the same as at $k-1$,
except each linear segment counts for two steps.) Whenever the system
size $N$ increases by a factor of 4, the diameter $d$ increases
by a factor of $\sim2$, leading to Eq.~(\ref{eq:3dia}). }
\label{fig:Koch3hanoi} 
\end{figure}

\begin{figure}

\includegraphics[clip,scale=0.3]{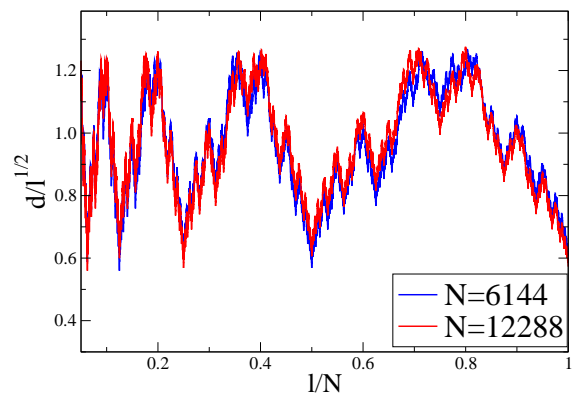}
\caption{Plot of the shortest path length between the origin
of HN3 and the $l$th site on two networks of extend $N=3\:2^{11}$
and $N=3\:2^{12}$. In both sets of data, we plot the path-distance
relative to the root of the separation between site $l$ and the origin
$(l=0)$ along the linear backbone. Then, all rescaled distances fluctuate
around a constant mean. Those fluctuations are very fractal, their
self-similarity becoming apparent when super-imposing the data for
both sizes $N$ on a relative distance scale with $l/N$. }
\label{fig:3dia}
\end{figure}

\begin{figure}
\includegraphics[clip,scale=0.3]{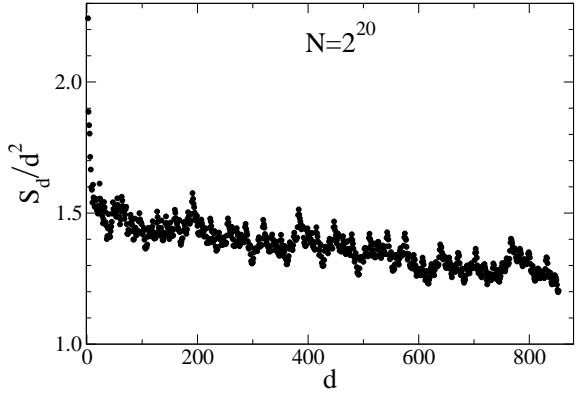} 
\caption{Plot of the number of sites (or {}``neighborhood'')
$S_{d}$ that can be reached from a given site within less than $d$
jumps on HN3. Averaged over many starting sites, $S_{d}/d^{2}$ slowly
converges to a constant, demonstrating that $S_{d}$ grows quadratically
with $d$. Note that some features due to the hierarchical structure
remain even after averaging over sites, such as the peaks at $d=192,$
384, 768, etc.}
\label{fig:HN3neighbors} 
\end{figure}

For HN3, it is simple to determine geometric properties, for instance,
its diameter $d$, which is the longest of the shortest paths between
any two sites, to wit, on a finite graph of size $N=N_{k}=2^{k}$
for $k\to\infty$ . Clearly, $d$ in this case would be the end-to-end
distance between sites $n=0$ and $n=N$ with the smallest number
of hops. Using a sequence of networks for $k=2,4,6,\ldots$, the diameter-path
looks like a Koch curve, see Fig.~\ref{fig:Koch3hanoi}. We can define
the path $\Pi_{k}$ as a sequence of jumps reaching from one end of
the $N=2^{k}$-long graph to the other via 
\begin{eqnarray}
\Pi_{0} & = & 1,\nonumber \\
\Pi_{2} & = & 1-2-1=\Pi_{0}-2-\Pi_{0},\nonumber \\
\Pi_{4} & = & 1-2-1-8-1-2-1=\Pi_{2}-8-\Pi_{2},\nonumber \\
 & \dots & , \\
\Pi_{k+2} & = & \Pi_{k}-2^{k+1}-\Pi_{k},\nonumber
\label{eq:recurPath3}
\end{eqnarray}
with an obvious notation of using {}``-'' to string together a sequence
of ever more complex moves. (While somewhat redundant here, introducing
this notation will prove useful for HN4 below.) Hence, the length
$d_{k}$ of each path $\Pi_{k}$ is given by 
\begin{eqnarray}
d_{k+2} & = & 2d_{k}+1\qquad {\rm for}\qquad N_{k+2}=4N_{k},
\label{eq:recurDia3}
\end{eqnarray}
thus, 
\begin{equation}
d\sim\sqrt{N}.
\label{eq:3dia}
\end{equation}
This property is demonstrated also in Fig.~\ref{fig:3dia}. In some
ways, this property is reminiscent of a square-lattice consisting of
$N$ lattice sites. The diameter (=diagonal) of this square is also
$\sim\sqrt{N}$. In this sense, presuming that each link corresponds to
a unit distance, HN3 is a fractal lattice of dimension 2, i. e.
filling the plane. As shown in Fig.~\ref{fig:HN3neighbors}, we also
find that the number of sites that can be reached from a given site
grows quadratically with the number of jumps allowed. It should be
noted, though, that we are employing the one-dimensional lattice
backbone as our metric to measure distances in the random walks in
Sec.~\ref{sec:Random-Walks}, which enforces a trivial fractal
dimension of $d_{f}=1$. We conclude that, while interesting in its own
right, HN3 is far from any mean-field behavior for which we would
expect typical distances to depend only on (some power of) $\ln N$.


\subsection{Distance Measure on HN4\label{sub:Distance-Measure-on}}

The situation is more interesting for the HN4. Using the notation
from Eq.~(\ref{eq:recurPath3}), we have
\begin{eqnarray}
\Pi_{0} & = & 1,\nonumber \\
\Pi_{1} & = & 1-1,\nonumber\\
\Pi_{2} & = & 1-2-1=\Pi_{0}-2-\Pi_{0},\nonumber \\
\Pi_{3} & = & 1-2-2-2-1=\Pi_{0}-3\times2-\Pi_{0},\\
\Pi_{4} & = & 1-1-4-4-4-1-1=\Pi_{1}-3\times4-\Pi_{1},\nonumber \\
\Pi_{5} & = & 1-2-1-8-8-8-1-2-1\nonumber\\
&=&\Pi_{2}-3\times8-\Pi_{2},\nonumber \\
\Pi_{6} & = & 1-2-1-7\times8-1-2-1\nonumber\\
&=&\Pi_{2}-7\times8-\Pi_{2},\nonumber
\label{eq:Path4}
\end{eqnarray}
and so on. Due to degeneracies at each level (which we have not listed),
one has to proceed to many more levels in the hierarchy to discern
the relevant pattern. In fact, any pattern evolves for an increasing
number of levels before it gets taken over by a new one, with two
patterns creating degeneracies at the crossover. Finally, we get (setting
the degeneracies aside)
\begin{eqnarray}
\Pi_{k} & = & \begin{cases}
\Pi_{k-2}-1\times2^{k-1}-\Pi_{k-2}, & (k=2),\\
\Pi_{k-3}-3\times2^{k-2}-\Pi_{k-3}, & (2<k\leq5),\\
\Pi_{k-4}-7\times2^{k-3}-\Pi_{k-4}, & (5<k\leq9),\\
\Pi_{k-5}-15\times2^{k-4}-\Pi_{k-5}, & (9<k\leq14),\\
\ldots,\end{cases}
\label{eq:recurPath4}
\end{eqnarray}
and so on. 

Note that the paths here do \emph{not} search out the longest possible
jump, as in Eq.~(\ref{eq:recurPath3}). Instead, the paths reach quickly
to some intermediate level and follow \emph{consecutive} jumps at
that level before trailing off in the end. As we will see repeatedly,
this is the main distinguishing feature discriminating between HN3
and HN4: Once a level is reached in the HN4, the entire graph can
be traversed at \emph{that} level, while in the HN3 any transport
\emph{must} climb down to lower levels (or merely jump back on that
level), see Figs.~\ref{fig:3hanoi}-\ref{fig:4hanoi}.

\begin{figure}
\includegraphics[clip,scale=0.3]{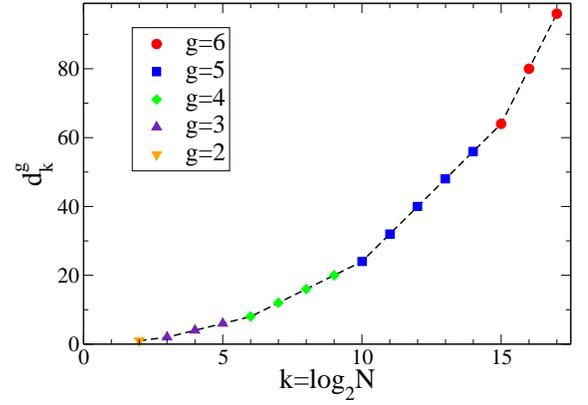}
\caption{Plot of the shortest end-to-end path length for HN4
networks of increasing backbone sizes $N=2^{k}$, obtained by simulation.
Note the piecewise-linear shape of the graph, which is reflected in
Eq.~(\ref{eq:dkgrecur}), for example. In turn, Eq.~(\ref{eq:crossover})
concerns only the {}``bends'' between each consecutive linear segment
$g$.}
\label{fig:dgk}
\end{figure}

\begin{figure}
\includegraphics[clip,scale=0.3]{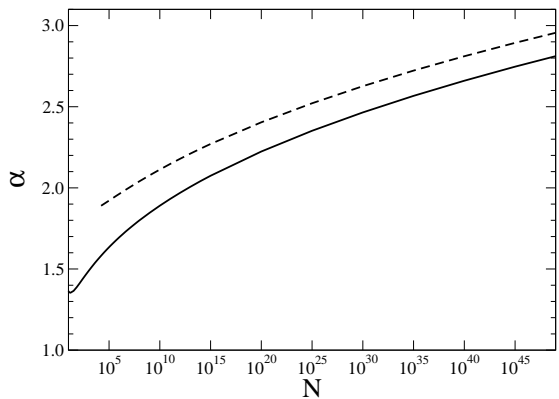}
\caption{Plot of the system-size dependence of the exponent
$\alpha=\alpha(N)$ defined in Eq.~(\ref{eq:alpha}). The solid curve
is the exact value based on Eq.~(\ref{eq:dkgsol}), and the dashed
curve is the asymptotic approximation also given in Eq.~(\ref{eq:alpha}).
This approximation is always an upper bound. Note that $\alpha(N)$
barely varies within a factor of  2 over 50 orders of magnitude.}
\label{fig:4dia}
\end{figure}

Corresponding to Eq.~(\ref{eq:recurPath4}), we obtain for the end-to-end
shortest paths (=diameters $d_{k}$)
\begin{eqnarray}
d_{k} & = & \begin{cases}
2d_{0}+1, & (k=2),\\
2d_{k-3}+3, & (2<k\leq5),\\
2d_{k-4}+7, & (5<k\leq9),\\
\ldots,\end{cases}
\label{eq:dkrecur}
\end{eqnarray}
which we can generalize into a single statement introducing a {}``generation''
index $g\geq2$,
\begin{eqnarray}
d_{k}^{g} & = &
2d_{k-g}^{g-1}+\left(2^{g-1}-1\right),\qquad\left(l_{g-1}<k\leq
l_{g}\right),
\label{eq:dkgrecur}
\end{eqnarray}
defining $l_{1}=1$, where in general
\begin{eqnarray}
l_{g} & = & l_{g-1}+g,\qquad l_{2}=2,
\label{eq:crossover}
\end{eqnarray}
demarcates the crossover point between the generations, see Fig.~\ref{fig:dgk}.
Eq.~(\ref{eq:crossover}) easily yields
\begin{eqnarray}
l_{g} & = & \frac{1}{2}g\left(g+1\right)-1\qquad\left(g\geq2\right).
\label{eq:crossoversol}
\end{eqnarray}
To obtain the asymptotic behavior for $d_{k}$, instead of solving
Eq.~(\ref{eq:dkgrecur}) for all $k$, we note that exactly on the
crossover points $k=l_{g}$ (i. e., $k-g=l_{g-1}$) we have
\begin{eqnarray}
d_{l_{g}}^{g} & = & 2d_{l_{g-1}}^{g-1}+\left(2^{g-1}-1\right).
\label{eq:crossoverrecur}
\end{eqnarray}
Defining $e_{g}=2^{g}d_{l_{g}}^{g}$, we get 
\begin{eqnarray}
e_{g} & = & e_{g-1}+\frac{1}{2}-2^{-g},
\end{eqnarray}
which is easily summed up to give
\begin{eqnarray}
d_{l_{g}}^{g} & = & \left(g-1\right)2^{g-1}+1.
\label{eq:dkgsol}
\end{eqnarray}
Remembering that $k=l_{g}$ and, from Eq.~(\ref{eq:crossoversol}),
that $g\sim\sqrt{2l_{g}}\sim\sqrt{2k}\sim\sqrt{\log_{2}N^{2}}$, we
finally get
\begin{eqnarray}
d_{k} & \sim &\frac{1}{2}
\sqrt{\log_{2}N^{2}}\,\,2^{\sqrt{\log_{2}N^{2}}}\qquad\left(N\to\infty\right)
\label{eq:dkasymp}
\end{eqnarray}
for the diameter of the HN4. As we would expect that the diameter
(or rather, the average of shortest paths) in a small-world graph
should behave as $d\sim\log N$, it is instructive to rewrite
Eq.~(\ref{eq:dkasymp}) as
\begin{eqnarray}
d_{k} & \sim & \left(\log_{2}N\right)^{\alpha}\quad{\rm with}
\quad\alpha\sim\frac{\sqrt{2\log_{2}N^{}}}{\log_{2}\log_{2}N}+\frac{1}{2}.
\label{eq:alpha}
\end{eqnarray}
Technically, of course, $\alpha$ diverges with $N$ and the diameter
grows faster than any power of $\log_{2}N$ but less than any however-small
power of $N$, unlike Eq.~(\ref{eq:3dia}). In reality, though, $\alpha$
varies only very slowly with $N$, ranging merely from $\alpha\approx1.5$
to 3 over \emph{fifty} decades, see Fig.~\ref{fig:4dia}.

\section{Random Walks\label{sec:Random-Walks}}

In the following we will study random walks on HN3 and HN4. For
simplicity, we have focused in our simulations only on two elementary
observables, the mean displacement with time, $\left\langle
\left|l\right|\right\rangle \sim t^{1/d_{w}}$, and the first-return
time distribution $Q(\Delta t)\sim\Delta t^{-\mu}$.  All walks are
controlled by one parameter, $p$, which determines the probability of
a walker to step off the lattice into the direction of a long-range
jump. In particular, on HN3 a walker steps off the lattice with
probability $p$ and jumps either to the left or right neighbor with
probability $\left(1-p\right)/2$, whereas on HN4 long-range jumps to
either the left or right occur with probability $p/2$ instead.  In
both cases, we should return to a simple one-dimensional
nearest-neighbor walk for $p\to0$, where $d_{w}=2$ and $\mu=3/2$,
although we would expect that limit to be singular. Here, we only
consider uniform values of $p$, independent of the sites (or level of
hierarchy). With a bit more algebra, the following considerations
could be extended to values that for each site depend of the level of
hierarchy, $p=p(i)$, for instance.

\subsection{Renormalization Group for Random Walks}

Here, we analyze the spatio-temporal rescaling of simple random walks
with nearest-neighbor jumps along the available links using the renormalization
group. First, we review the method using as a simple example the well-known
one-dimensional walk. Walks on such a graph obey simple diffusion,
$d_{w}=2$, which implies that each rescaling of space entails a rescaling
of time according to
\begin{eqnarray}
N\to N'=2N & \qquad & T\to T'=2^{d_{w}}T=4T.
\label{eq:1dWalkscale}
\end{eqnarray}
This is synonymous with the asymptotic form with the mean-square displacement
\begin{equation}
\left\langle \left|l\right|\right\rangle \sim t^{1/d_{w}},
\label{MSDeq}
\end{equation}
 which defines the anomalous dimension of the walk in terms of the
exponent $d_{w}$. While the result for the one-dimensional walk can
be easily obtained with much simpler means, it serves as a pedagogical
example of calculating first-passage and first-return times using
RG on more complex structures. Later on, reference to this presentation
will allow us to avoid excessive algebra.

\subsubsection{RG for the 1d random walk\label{sub:RG-for-2RW}}

As a pedagogical example, we present here the theory as it will be
applied in Sec. \ref{sub:RG-for-3RW}. We consider a biased random
walk on a finite one-dimensional line. The master equations for such
a random walk on a lattice of length $N=2^{K+1}$ with reflecting
boundaries are given by

\begin{eqnarray}
P_{0,t+1} & = & p\, P_{1,t},\nonumber \\
P_{1,t+1} & = & P_{0,t}+p\, P_{2,t},\nonumber \\
P_{l,t+1} & = & (1-p)\, P_{l-1,t}+p\, P_{l+1,t}\qquad(2\leq l\leq
N-2),\nonumber \\
P_{N-1,t+1} & = & (1-p)\, P_{N-2,t}+P_{N,t},\nonumber \\
P_{N,t+1} & = & (1-p)\, P_{N-1,t},
\label{eq:1dRW}
\end{eqnarray}
where $P_{l,t}$ denotes the probability of a walker to be at site
$l$ at time $t$, $p$ is the probability expressing the biasing
for left or right hops. Since we want to start the walks at time $t=0$
at the origin $l=0$, these equations have the initial condition
\begin{eqnarray}
P_{l,0} & = & \delta_{l,0}.
\label{eq:1dRWinit}
\end{eqnarray}

To facilitate renormalization this non-equilibrium process, we introduce
a generating function
\begin{eqnarray}
\tilde{P}_{l}(z) & = & \sum_{t=0}^{\infty}P_{l,t}z^{t}
\label{eq:generator}
\end{eqnarray}
for all $0\leq l\leq N$. Incorporating the initial condition in
Eq.~(\ref{eq:1dRWinit}), Eqs.~(\ref{eq:1dRW}) transform into
\begin{eqnarray}
\tilde{P}_{0} & = & a\,\tilde{P}_{1}+1,\nonumber \\
\tilde{P}_{1} & = & c\,\tilde{P}_{0}+a\,\tilde{P}_{2},\nonumber \\
\tilde{P}_{l} & = & b\,\tilde{P}_{l-1}+a\,\tilde{P}_{l+1}\qquad(2\leq l\leq N-2),\nonumber \\
\tilde{P}_{N-1} & = & b\,\tilde{P}_{N-2}+d\,\tilde{P}_{N},\nonumber \\
\tilde{P}_{N} & = & b\,\tilde{P}_{N-1},\label{eq:1dRWgen}
\end{eqnarray}
where we have inserted generalized {}``hopping rates'' in preparation
for the RG. Initially, at the $k=0$th RG step, it is
\begin{eqnarray}
a^{(0)} & = & p\, z,\nonumber \\
b^{(0)} & = & (1-p)\, z,\label{eq:RGinitpara}\\
c^{(0)} & = & z,\nonumber \\
d^{(0)} & = & z,\nonumber 
\end{eqnarray}
which provides a sufficient number of renormalizable parameters that
are potentially required to consider special sites at both boundaries. 

A single step of applying the RG consists of solving Eqs.~(\ref{eq:1dRWgen})
for $\tilde{P_{l}}$ with odd values of $l$ (which is trivial here,
as they are already expressed explicitly in terms of even ones) and
eliminating them from the equations for the even $l$. After that
elimination, we can rewrite the equations for even $l$ as
\begin{eqnarray}
\tilde{P}_{0} & = & \frac{a^{2}}{1-ac}\,\tilde{P}_{2}+\frac{1}{1-ac},\nonumber \\
\tilde{P}_{2} & = & \frac{bc}{1-2ab}\,\tilde{P}_{0}+\frac{a^{2}}{1-2ab}\,\tilde{P}_{4},\nonumber \\
\tilde{P}_{2l} & = & \frac{b^{2}}{1-2ab}\,\tilde{P}_{2l-2}+\frac{a^{2}}{1-2ab}\,\tilde{P}_{2l+2}\quad\left(2\leq l\leq\frac{N}{2}-2\right),\nonumber \\
\tilde{P}_{N-2} & = & \frac{b^{2}}{1-2ab}\,\tilde{P}_{N-4}+\frac{ad}{1-2ab}\,\tilde{P}_{N},\nonumber \\
\tilde{P}_{N} & = & \frac{b^{2}}{1-bd}\,\tilde{P}_{N-2}.\label{eq:1dRWgen-reno}
\end{eqnarray}
Comparing these equations with Eqs.~(\ref{eq:1dRWgen}) allows to
extract the RG recursion equations. {[}Note that superscripts referring
the $k$th RG step have been suppressed thus far in Eqs.~(\ref{eq:1dRWgen})
and~(\ref{eq:1dRWgen-reno}).]

Before we analyze the first return time at the boundary specifically,
we can use the equation for bulk sites $l$ in~(\ref{eq:1dRWgen-reno})
to extract already the diffusion exponent $d_{w}$. A comparison of
the respective coefficients in Eqs.~(\ref{eq:1dRWgen}) and~(\ref{eq:1dRWgen-reno})
yields
\begin{eqnarray}
a^{(k+1)}&=&\frac{\left(a^{(k)}\right)^{2}}{1-2a^{(k)}b^{(k)}},\nonumber\\ 
b^{(k+1)}&=&\frac{\left(b^{(k)}\right)^{2}}{1-2a^{(k)}b^{(k)}}.
\label{eq:RGbulk}
\end{eqnarray}
These recursions converge for $k\to\infty$ towards fixed points
$\left(a^{*},b^{*}\right)$ that characterize the dynamics in the
infinite-time limit (which corresponds to the limit of $z\to1^{-}$).
The trivial fixed point $a^{*}=b^{*}=0$ is unphysical, as it can
not be reached from the initial conditions in (\ref{eq:RGinitpara})
for any choice of $p$ (and $z=1$). The physical fixed points are
$\left(a^{*},b^{*}\right)=\left(1,0\right)$, which is reached for
any bias $p>\frac{1}{2}$, or $\left(a^{*},b^{*}\right)=\left(0,1\right)$,
reached for $p<\frac{1}{2}$; finally, $\left(a^{*},b^{*}\right)=\left(\frac{1}{2},\frac{1}{2}\right)$
can only be reached by entirely unbiased walks for $p=\frac{1}{2}$.
To explore the behavior for large but finite times, we expand the
RG recursions in (\ref{eq:RGbulk}) to first order in $\epsilon=1-z$
by writing for $y\in\left\{ a,b\right\} $:
\begin{eqnarray}
y^{(k)} & \sim & y^{*}+y_{1}^{(k)}\epsilon+\ldots.
\label{eq:epsilon1}
\end{eqnarray}
Inserting the Ansatz in Eq.~(\ref{eq:epsilon1}) into the recursions
in Eqs.~(\ref{eq:RGbulk}), we obtain near the fixed point with
$a^{*}=b^{*}=\frac{1}{2}$:
\begin{eqnarray}
a_{1}^{(k+1)} & = & 3a_{1}^{(k)}+b_{1}^{(k)},\nonumber \\
b_{1}^{(k+1)} & = & a_{1}^{(k)}+3b_{1}^{(k)},
\label{eq:1dRWunbias}
\end{eqnarray}
with the result that 
\begin{eqnarray*}
a_{1}^{(k)} & = & b_{1}^{(k)}\propto4^{k}.
\end{eqnarray*}
This implies that as space rescales by a factor of 2 (i. e.,
eliminating all odd-index sites), time rescales by a factor of 4, as
indicated in Eq.~(\ref{eq:1dWalkscale}) for an unbiased random walk,
leading to $d_{w}=2$. The same analysis for either of the biased fixed
points yields that, for example, $a^{(k)}\equiv0$ beyond any power of
$\epsilon$ and, with the Ansatz $b^{(k)}\sim1+b_{1}^{(k)}\epsilon$ in
Eqs.~(\ref{eq:RGbulk}), $b_{1}^{(k)}\propto2^{k}$. Following the
interpretation in Eq.~(\ref{eq:1dWalkscale}), this would imply
$d_{w}=1$ and we find the familiar result that with the slightest
bias, i. e., $p<\frac{1}{2}$ or $p>\frac{1}{2}$, the motion at large
length and time scales is dominated by the constant-velocity drift
upon reaching the bulk.

In this scenario of a bias, average first-return times are clearly
system-size independent constants: A walker with a bias \emph{towards}
the origin ($p>\frac{1}{2}$) will drift back recurrently after only
small excursions; a walker with a bias \emph{away} from the origin
($p<\frac{1}{2}$) returns at most a finite number of times in short
order until the drift eventually carries it away without further recurrence.
In the following, we therefore focus exclusively on the unbiased case
$p=\frac{1}{2}$. Then, we can equate $a=b$ at every step, to get
from Eq.~(\ref{eq:RGbulk}): 
\begin{eqnarray}
a^{(k+1)}=\frac{\left(a^{(k)}\right)^{2}}{1-2\left(a^{(k)}\right)^{2}}.
\label{eq:RGbulkunbias}
\end{eqnarray}

To derive the return-time behavior, we have to examine
Eqs.~(\ref{eq:1dRWgen-reno}) more closely. Comparing coefficients also
in the boundary terms leads to
\begin{eqnarray}
c^{(k+1)}&=&\frac{a^{(k)}c^{(k)}}{1-2\left(a^{(k)}\right)^{2}},\nonumber\\
d^{(k+1)}&=&\frac{a^{(k)}d^{(k)}}{1-2\left(a^{(k)}\right)^{2}}.
\label{eq:RGupper}
\end{eqnarray}
For large $k$, both $c^{(k)}$ and $d^{(k)}$ are entrained with
$a^{(k)}$, and we obtain a consistent and closed set of relations
for all coefficients in Eqs.~(\ref{eq:1dRWgen}) and (\ref{eq:1dRWgen-reno})
by identifying $c=d=2a$. Further renormalizing 
\begin{eqnarray}
\tilde{P}_{l}^{(k+1)} & = &
\left[1-2\left(a^{(k)}\right)^{2}\right]\,\tilde{P}_{2l}^{(k)}
\label{eq:amplitude-reno}
\end{eqnarray}
 ensures invariance of the constant term at the lower boundary that
originated from the unit initial condition in Eq.~(\ref{eq:1dRWinit}).

After $k=K$ RG steps, the system has reduced to 
\begin{eqnarray}
\tilde{P}_{0}^{(K)} & = & a^{(K)}\tilde{P}_{1}^{(K)}+1,\nonumber \\
\tilde{P}_{1}^{(K)} & = & 2a^{(K)}\tilde{P}_{0}^{(K)}+2a^{(K)}\tilde{P}_{2}^{(K)},\label{eq:1d-elementary}\\
\tilde{P}_{2}^{(K)} & = & a^{(K)}\tilde{P}_{1}^{(K)},\nonumber 
\end{eqnarray}
which yields 
\begin{eqnarray}
\tilde{P}_{0}^{(K)} & = &
\frac{1-2\left(a^{(K)}\right)^{2}}{1-4\left(a^{(K)}\right)^{2}}.
\label{eq:P0K1d}
\end{eqnarray}
Using Eq.~(\ref{eq:amplitude-reno}) in turn obtains
\begin{eqnarray}
\tilde{P}_{0}^{(0)} & = &
\frac{\tilde{P}_{0}^{(K)}}{\prod_{k=0}^{K-1}\left[1-2\left(a^{(k)}\right)^{2}\right]}\nonumber\\
&=&\frac{1-2\left(a^{(K)}\right)^{2}}{1-4\left(a^{(K)}\right)^{2}}\:\prod_{k=0}^{K-1}\frac{1}{\left[1-2\left(a^{(k)}\right)^{2}\right]}.
\label{eq:P001d}
\end{eqnarray}
It is a well-known fact that the generating functions for being at
the origin, $\tilde{P}_{0}^{(0)}$, and for the first-return probability
to the same site, $\tilde{Q}_{0}$, satisfy the following simple
relation \cite{Redner01}:
\begin{eqnarray}
\tilde{Q}_{l} & = & 1-\frac{1}{\tilde{P}_{l}^{(0)}}.
\label{eq:Q}
\end{eqnarray}
Note that a recurrent walk (with $\tilde{Q}_{0}=1$) requires that
$ $ $\tilde{P}_{0}^{(0)}$ diverges at long times (i. e. $z\to1^{-}$).
In our one-dimensional walk here, it is $a^{(K)}\to a^{*}=\frac{1}{2}$,
on behalf of which the denominator of $\tilde{P}_{0}^{(0)}$ in Eq.
(\ref{eq:P001d}) has a zero, making the walk recurrent. In more detail,
it is 
\begin{eqnarray}
\tilde{Q}_{0} & = &
1-\frac{1-4\left(a^{(K)}\right)^{2}}{1-2\left(a^{(K)}\right)^{2}}\:\prod_{k=0}^{K-1}\left[1-2\left(a^{(k)}\right)^{2}\right].
\label{eq:Q1d}
\end{eqnarray}

We intend to extract the exponent $\mu$ for the first-return probability
distribution, which on a finite but large system of size $N\to\infty$
behaves as
\begin{eqnarray}
Q_{0}(t) & \sim & t^{-\mu}e^{-t/\tau_{N}}\qquad(t\to\infty),
\label{eq:Qt}
\end{eqnarray}
where $\tau_{N}$ is a cut-off time scale that diverges in some form
with $N$. As the physics of the return probabilities at any finite
time should not change if the system size becomes infinite independently,
it must be $\mu>1$ for $Q_{0}(t)$ to remain normalizable. Based
on that observation \cite{Redner01}, we need to calculate the first
two moments of $Q_{0}(t)$, corresponding to an expansion of $\tilde{Q}_{0}(z)$
to 2nd order in $\epsilon=1-z$, to extract $\mu$. Since $\mu>1$,
the normalization integral 
\begin{eqnarray}
{\cal N} & \sim & \int^{\infty}dt\, t^{-\mu}e^{-t/\tau_{N}}\sim O(1)
\label{eq:taunorm}
\end{eqnarray}
is dominated by its behavior for small $t$, which is irrelevant in
detail except for the fact that it makes the integral become a constant
independent of $\tau_{N}$. If we further assume that $\mu<2$, then
for all $m\geq1$, the integrals for those $m$th moments \emph{do}
diverge with $\tau_{N}$, 
\begin{eqnarray}
\left\langle t^{m}\right\rangle _{N} & \sim\frac{1}{{\cal N}} &
\int_{0}^{\infty}dt\, t^{m-\mu}e^{-t/\tau_{N}}\sim\tau_{N}^{m+1-\mu}.
\label{eq:taumoment}
\end{eqnarray}
From the ratio of $\left\langle t\right\rangle _{N}$ and $\left\langle t^{2}\right\rangle _{N}$,
we obtain then
\begin{eqnarray}
\mu & =\lim_{N\to\infty} & 2+\frac{1}{1-\frac{\left\langle
    t^{2}\right\rangle _{N}}{\left\langle t\right\rangle _{N}}}.
\label{eq:mu}
\end{eqnarray}

Luckily, due to the leading zero in $1-4\left(a^{(K)}\right)^{2}$,
any other factor in Eq.~(\ref{eq:Q1d}) only needs to be expanded
to first order in $\epsilon$. Extending the Ansatz in Eq.~(\ref{eq:epsilon1})
for $a^{(k)}$ to 2nd order, we obtain here:
\begin{eqnarray*}
a^{(K)} & \sim & \frac{1}{2}-\frac{1}{2}\times4^{K}\epsilon+5\times16^{K}\epsilon^{2}+\ldots,\\
1-4\left(a^{(K)}\right)^{2} & \sim & 8\times4^{K}\epsilon+116\times16^{K}\epsilon^{2}+\ldots,\\
1-2\left(a^{(K)}\right)^{2} & \sim & \frac{1}{2}+4\times4^{K}\epsilon+\ldots,\\
\prod_{k=0}^{K-1}\left[1-2\left(a^{(k)}\right)^{2}\right] & \sim & \prod_{k=0}^{K-1}\left[\frac{1}{2}+4\times4^{k}\epsilon+\ldots\right],\\
 & \sim & 2^{-K}\left[1+8\epsilon\sum_{k=0}^{K-1}4^{k}+\ldots\right],\\
 & \sim & 2^{-K}+\frac{8}{3}\,2^{K}\epsilon+\ldots.
\end{eqnarray*}
Inserting these expressions into Eq.~(\ref{eq:Q1d}), we get
\begin{eqnarray}
\tilde{Q}_{0} & \sim &
1-16\times2^{K}\epsilon+\frac{952}{3}\times8^{K}\epsilon^{2}+\ldots.
\label{eq:Q1dasymp}
\end{eqnarray}
The moments of $Q_{0}(t)$ are obtained via derivatives of $\tilde{Q}_{0}(z)$,
i.~e. 
\begin{eqnarray}
\left\langle t^{m}\right\rangle  & = &
\left[z\,\frac{d}{dz}\right]^{m}\tilde{Q}_{0}(z)\vert_{z=1}.
\label{eq:def-moments}
\end{eqnarray}
Applied to Eq.~(\ref{eq:Q1dasymp}), we calculate for the dominant
asymptotics of the moments:
\begin{eqnarray*}
\left\langle t\right\rangle _{K} & \sim & 2^{K}\sim N,\\
\left\langle t^{2}\right\rangle _{K} & \sim & 8^{K}\sim N^{3},
\end{eqnarray*}
which from Eq.~(\ref{eq:mu}) leads to the familiar first-return exponent
of a one-dimensional walk, 
\begin{eqnarray}
\mu & = & \frac{3}{2}.
\label{eq:1dmu}
\end{eqnarray}

\subsubsection{RG for the random walk on HN3\label{sub:RG-for-3RW}}

We follow the discussion for one-dimensional walkers above to model
diffusion on HN3. The lesson of the previous section is that we only
need to consider the bulk equations to extract the diffusion exponent
$d_{w}$, and supplement with the calculation on the final system
at the end of the $k=K$th step of the RG to obtain the first-return
exponent for a system of size $N=2^{K+1}$. We can section the line
into segments centered around sites $l$ with $i=1$ in Eq.~(\ref{eq:numbering}),
i. e. $n=2(2j+1)$. Such a site $l=n$ is surrounded by two sites
of odd index, which are \emph{mutually} linked. Furthermore, $n$
is linked by a long-distance jump to a site also of type $i=1$ at
$l=n\pm4$ in the neighboring segment, where the direction does not
matter here. The sites $l=n\pm2$, which are shared at the boundary
between adjacent segments also have even index, but their value of
$i\geq2$ is undetermined and irrelevant for the immediate RG step,
as they have a long-distance jump to some sites $l=m_{\pm}$ at least
eight sites away. 

In each segment in the bulk, the master-equation reads 
\begin{eqnarray}
P_{n+2,t+1} & = & \frac{1-p}{2}\left[P_{n+3,t}+P_{n+1,t}\right]+p\, P_{m_{+},t},\nonumber \\
P_{n+1,t+1} & = & \frac{1-p}{2}\left[P_{n+2,t}+P_{n,t}\right]+p\, P_{n-1,t},\nonumber \\
P_{n,t+1} & = & \frac{1-p}{2}\left[P_{n+1,t}+P_{n-1,t}\right]+p\, P_{n\pm4,t},\label{eq:3RW}\\
P_{n-1,t+1} & = & \frac{1-p}{2}\left[P_{n,t}+P_{n-2,t}\right]+p\, P_{n+1,t},\nonumber\\
P_{n-2,t+1} & = & \frac{1-p}{2}\left[P_{n-1,t}+P_{n-3,t}\right]+p\,
P_{m_{-},t}.\nonumber
\end{eqnarray}
Here, $p$ is the (uniform) probability for a walker to take a long-range
jump, whereas $(1-p)/2$ is the probability to jump either left or
right towards a nearest-neighbor site along the backbone. Without
restriction of generality, let us assume that we happened to let all
walks start from a site within this segment. Note that, unlike for
the $1d$-walk above, there are \emph{three} distinct types of sites
even in the bulk of this problem: $P_{n\pm2,t}$, $P_{n\pm1,t}$,
and $P_{n,t}$. In principle, one could expect three distinct return-time
behaviors as a result. We will demonstrate below that, indeed, different
sites possess differences in their return-time behavior, depending
on their level in the hierarchy, but all scale with the same exponent
$\mu$. Here, we choose to have the walk start on site $n-2$ within
this segment:
\begin{eqnarray}
P_{l,0} & = & \delta_{l,n-2}.
\label{eq:3RWinit}
\end{eqnarray}

\begin{figure}
\includegraphics[bb=0bp 550bp 380bp 750bp,clip,scale=0.6]{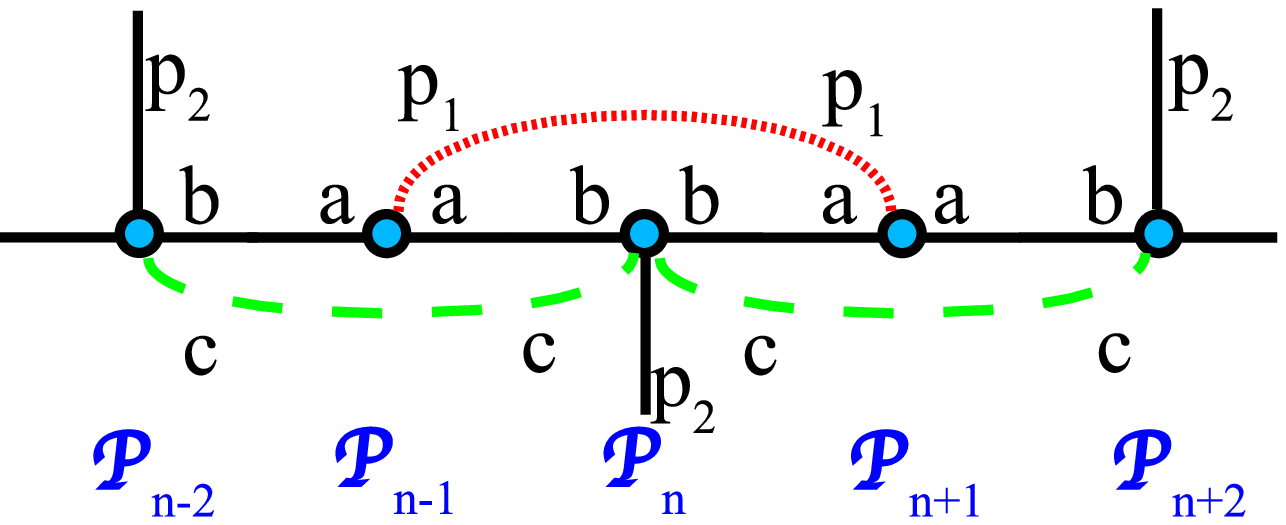}
\includegraphics[bb=0bp 550bp 320bp 750bp,clip,scale=0.6]{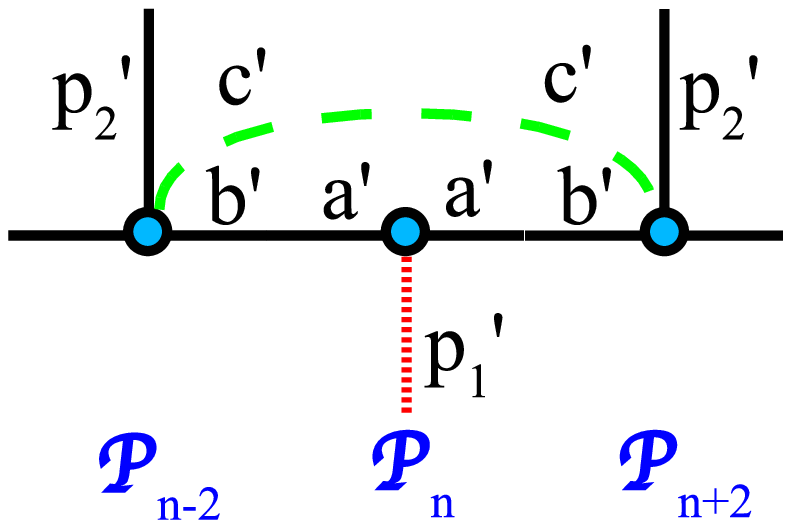} 

\caption{Depiction of the (exact) RG step for random walks on
  HN3. Hopping rates from one site to another along a link are labeled
  at the originating site. The RG step consists of tracing out
  odd-labeled variables $\tilde{P}_{n\pm1}$ in the top graph and
  expressing the renormalized rates $(a',b',c',p_{1}',p_{2}')$ on the
  bottom in terms of the ones $(a,b,c,p_{1},p_{2})$ from the
  top. The node $\tilde{P}_{n}$, bridged by a (dotted) link between
  $\tilde{P}_{n-1}$ and $\tilde{P}_{n+1}$, is special as it
  \emph{must} have $n=2(2j+1)$ and is to be decimated at the following
  RG step, justifying the designation of $p_{1}'$. Note that the
  original graph in Fig.~\ref{fig:3hanoi} does not have the green,
  dashed links with hopping rates $(c,c')$, which \emph{emerge} during
  the RG recursion. }
\label{fig:RG3RW} 
\end{figure}

Using the generating function in Eq.~(\ref{eq:generator}) on
Eqs.~(\ref{eq:3RW}-\ref{eq:3RWinit}) yields:
\begin{eqnarray}
\tilde{P}_{n+2} & = & a\left[\tilde{P}_{n+3}+\tilde{P}_{n+1}\right]+c\left[\tilde{P}_{n+4}+\tilde{P}_{n}\right]+p_{2}\,\tilde{P}_{m_{+}},\nonumber \\
\tilde{P}_{n+1} & = & b\left[\tilde{P}_{n+2}+\tilde{P}_{n}\right]+p_{1}\,\tilde{P}_{n-1},\nonumber \\
\tilde{P}_{n} & = & a\left[\tilde{P}_{n+1}+\tilde{P}_{n-1}\right]+c\left[\tilde{P}_{n+2}+\tilde{P}_{n-2}\right]+p_{2}\,\tilde{P}_{n\pm4},\nonumber\\
\tilde{P}_{n-1} & = & b\left[\tilde{P}_{n}+\tilde{P}_{n-2}\right]+p_{1}\,\tilde{P}_{n+1},\label{eq:3RWgen} \\
\tilde{P}_{n-2} & = &
a\left[\tilde{P}_{n-1}+\tilde{P}_{n-3}\right]+c\left[\tilde{P}_{n}+\tilde{P}_{n-4}\right]+p_{2}\,\tilde{P}_{m_{-}}+1,\nonumber 
\end{eqnarray}
where we have again absorbed the parameters $p$ and $z$ into general
{}``hoping rates'', which are initially
\begin{eqnarray}
a^{(0)} & = & b^{(0)}=\frac{z}{2}\left(1-p\right),\nonumber \\
c^{(0)} & = & 0,\label{eq:3parainit}\\
p_{1}^{(0)} & = & p_{2}^{(0)}=z\, p.\nonumber 
\end{eqnarray}
Note that we have added new terms with a parameter $c$, which is
zero initially. As Fig.~\ref{fig:RG3RW} depicts, such links are not
present in the original network HN3, but have to be taken into account
during the RG process. In all, a surprising number of parameters,
five in all, is required even for the most symmetric set of initial
conditions to obtain a closed set of RG recursion equations. Unlike
for the one-dimensional walk in Eq.~(\ref{eq:1dRWgen}), the parameters
$a$ and $b$ here do not express a directional bias or drift along
the backbone. Instead, they are necessary merely to distinguish between
hops out of (currently) odd and even sites, respectively, as shown
in Fig.~\ref{fig:RG3RW}.

The RG update step consist of eliminating from these five equations
those two that refer to an odd index, $n\pm1$. As a first step, adding
the two relations referring to the indices $n\pm1$, we obtain
\begin{eqnarray*}
\tilde{P}_{n+1}+\tilde{P}_{n-1} & = &
\frac{b}{1-p_{1}}\left[\tilde{P}_{n+2}+2\tilde{P}_{n}+\tilde{P}_{n-2}\right],
\end{eqnarray*}
which allows us to eliminate any reference to odd sites $n\pm1$ from
the middle relation in~(\ref{eq:3RWgen}). Furthermore, solving for
$\tilde{P}_{n\pm1}$ explicitly,
\begin{eqnarray*}
\tilde{P}_{n\pm1} & = &
\frac{b}{1-p_{1}^{2}}\,\tilde{P}_{n\pm2}+\frac{bp_{1}}{1-p_{1}^{2}}\,\tilde{P}_{n\mp2}+\frac{b}{1-p_{1}}\,\tilde{P}_{n},
\end{eqnarray*}
and inserting into the relations for $\tilde{P}_{n\pm2}$ in Eq.~(\ref{eq:3RWgen})
results in 
\begin{eqnarray}
\tilde{P}_{n+2} & = & \frac{\left[ab+c\left(1-p_{1}\right)\right]\left(1+p_{1}\right)}{1-p_{1}^{2}-2ab}\left[\tilde{P}_{n+4}+\tilde{P}_{n}\right]\nonumber \\
 &&\quad+\frac{abp_{1}}{1-p_{1}^{2}-2ab}\left[\tilde{P}_{n+6}+\tilde{P}_{n-2}\right]\nonumber \\
 &&\quad+\frac{p_{2}\left(1-p_{1}^{2}\right)}{1-p_{1}^{2}-2ab}\,\tilde{P}_{m_{+}},\nonumber\\
\tilde{P}_{n} & = & \frac{ab+c\left(1-p_{1}\right)}{1-p_{1}-2ab}\left[\tilde{P}_{n+2}+\tilde{P}_{n-2}\right]\nonumber \\
 &&\quad+\frac{p_{2}\left(1-p_{1}\right)}{1-p_{1}-2ab}\,\tilde{P}_{n\pm4},\label{eq:3RWgen_reno}\\
\tilde{P}_{n-2} & = & \frac{\left[ab+c\left(1-p_{1}\right)\right]\left(1+p_{1}\right)}{1-p_{1}^{2}-2ab}\left[\tilde{P}_{n}+\tilde{P}_{n-4}\right]\nonumber \\
 &&\quad+\frac{abp_{1}}{1-p_{1}^{2}-2ab}\left[\tilde{P}_{n+2}+\tilde{P}_{n-6}\right]\nonumber \\
 &  &
\quad+\frac{p_{2}\left(1-p_{1}^{2}\right)}{1-p_{1}^{2}-2ab}\,\tilde{P}_{m_{-}}+\frac{1-p_{1}^{2}}{1-p_{1}^{2}-2ab}.\nonumber
\end{eqnarray}
Given that $n$ is only once divisible by 2, either $n-2$ or $n+2$
must be divisible by 2 at most twice, and we assume without restriction
of generality that $n+2$ satisfies this description, whereas $n-2$
is divisible by a higher power of 2. (This also selects the upper
sign in the index {}``$n\pm4$''). As in Eq.~(\ref{eq:amplitude-reno}),
we now add the proper superscript for the $k$th RG step and obtain
the renormalized generating functions at step $k+1$ as
\begin{eqnarray}
\tilde{P}_{l}^{(k+1)} &
=\frac{1-\left(p_{1}^{(k)}\right)^{2}-2a^{(k)}b^{(k)}}{1-\left(p_{1}^{(k)}\right)^{2}}
& \tilde{P}_{2l}^{(k)}.
\label{eq:3RWampli-reno}
\end{eqnarray}
The proportionality factor in this relation arises from the necessity
to preserve the unity of the initial condition, as in Eq.~(\ref{eq:amplitude-reno})
above. Note that we would have obtained the \emph{identical} factor,
if we had chosen to start the walker from the central site $n$ instead
of $n+2$. Starting at a site like $n\pm1$, in turn, would have provided
a different factor, 
\begin{eqnarray}
\tilde{P}_{l}^{(k+1)} & = &
\frac{1-p_{1}^{(k)}-2a^{(k)}b^{(k)}}{1-p_{1}^{(k)}}\,\tilde{P}_{2l}^{(k)},
\label{eq:3RWampli-reno2}
\end{eqnarray}
potentially leading to a distinct return-time behavior. Yet, asymptotically
for large times and distances both alternatives prove identical to
sufficiently high order as to not affect the scaling discussed below. 

Proceeding with the RG step, the role of the central site in the new
segments is then specified via $\frac{n+2}{2}\to n$ such that $\tilde{P}_{n-2}^{(k)}\to\tilde{P}_{n-2}^{(k+1)}$,
$\tilde{P}_{n}^{(k)}\to\tilde{P}_{n-1}^{(k+1)}$, $\tilde{P}_{n+2}^{(k)}\to\tilde{P}_{n}^{(k+1)}$,
and correspondingly for the functions on the left-hand side of Eqs.~(\ref{eq:3RWgen_reno}), which now reads
\begin{eqnarray}
\tilde{P}_{n}^{(k+1)} & = & a^{(k+1)}\left[\tilde{P}_{n+1}^{(k+1)}+\tilde{P}_{n-1}^{(k+1)}\right]\nonumber \\
 &&\quad+c^{(k+1)}\left[\tilde{P}_{n+2}^{(k+1)}+\tilde{P}_{n-2}^{(k+1)}\right]+p_{2}^{(k+1)}\,\tilde{P}_{n\pm4}^{(k+1)},\nonumber \\
\tilde{P}_{n-1}^{(k+1)} & = & b^{(k+1)}\left[\tilde{P}_{n}^{(k+1)}+\tilde{P}_{n-2}^{(k+1)}\right]+p_{1}^{(k+1)}\,\tilde{P}_{n+1}^{(k+1)},\nonumber\\
\tilde{P}_{n-2}^{(k+1)} & = & a^{(k+1)}\left[\tilde{P}_{n-1}^{(k+1)}+\tilde{P}_{n-3}^{(k+1)}\right]\label{eq:3RWgen_reno_new} \\
 &&\quad+c^{(k+1)}\left[\tilde{P}_{n}^{(k+1)}+\tilde{P}_{n-4}^{(k+1)}\right]\nonumber \\
 &&\quad+p_{2}^{(k+1)}\,\tilde{P}_{m_{-}}^{(k+1)}+1,\nonumber
\end{eqnarray}
These equations have \emph{exactly} the desired form of the corresponding
unrenormalized ones in Eqs.~(\ref{eq:3RWgen}), necessitating renomalization
recursions for the parameters of the form 
\begin{eqnarray}
a^{(k+1)} & = & \frac{\left[a^{(k)}b^{(k)}+c^{(k)}\left(1-p_{1}^{(k)}\right)\right]\left(1+p_{1}^{(k)}\right)}{1-\left(p_{1}^{(k)}\right)^{2}-2a^{(k)}b^{(k)}},\nonumber \\
\nonumber \\b^{(k+1)} & = & \frac{a^{(k)}b^{(k)}+c^{(k)}\left(1-p_{1}^{(k)}\right)}{1-p_{1}^{(k)}-2a^{(k)}b^{(k)}},\nonumber \\
\nonumber \\c^{(k+1)} & = & \frac{a^{(k)}b^{(k)}p_{1}^{(k)}}{1-\left(p_{1}^{(k)}\right)^{2}-2a^{(k)}b^{(k)}},\label{eq:RG3RWfp}\\
\nonumber \\p_{1}^{(k+1)} & = & \frac{p_{2}^{(k)}\left(1-p_{1}^{(k)}\right)}{1-p_{1}^{(k)}-2a^{(k)}b^{(k)}},\nonumber \\
\nonumber \\p_{2}^{(k+1)} & = &
\frac{p_{2}^{(k)}\left[1-\left(p_{1}^{(k)}\right)^{2}\right]}{1-\left(p_{1}^{(k)}\right)^{2}-2a^{(k)}b^{(k)}}.\nonumber 
\end{eqnarray}
The analysis of the fixed points for $k\to\infty$ of Eqs.~(\ref{eq:RG3RWfp})
is surprisingly subtle. Of course, we obtain a rather simple fixed
point for the choice of $p=0$, which eliminates all long-range jumps.
Then, $c^{(k)}$, $p_{1}^{(k)}$, and $p_{2}^{(k)}$ vanish for $k=0$
in the initial conditions in~(\ref{eq:3parainit}) and remain zero
for all $k>0$, according to Eqs.~(\ref{eq:RG3RWfp}). As a consequence,
the distinction between $a^{(k)}$ and $b^{(k)}$ disappears and both
recursions reduce \emph{exactly} to that for the unbiased one-dimensional
walk in Eq.~(\ref{eq:RGbulkunbias}) with $a^{*}=b^{*}=\frac{1}{2}$,
leading to ordinary diffusion with $d_{w}=2$ and $\mu=\frac{3}{2}$,
as discussed in Sec.~\ref{sub:RG-for-2RW}. Clearly, this fixed point
is unstable with respect to variations in $p$.

For \emph{any} probability $p>0$ inserted in Eqs.~(\ref{eq:3parainit}),
the recursions in (\ref{eq:RG3RWfp}) evolve towards an apparent fixed
point at $a^{*}=b^{*}=c^{*}=0$ and $p_{1}^{*}=p_{2}^{*}=1$. But
these recursions are \emph{singular} at such a fixed point, requiring
a more detailed investigation. If we choose an arbitrarily small $\delta>0$
and set $p=1-\delta$, we find that for all $k\geq0$ it is $a^{(k)}\sim b^{(k)}\sim c^{(k)}=O(\delta)$
and $p_{1}^{(k)}\sim p_{2}^{(k)}=1-O(\delta)$. Hence, setting $\delta=0$
in the end indeed validates this fixed point. Yet, the physics of
this fixed point, corresponding to choosing $p=1$, is trivial and
does not reflect the numerical observations: Strictly for $p=1$ there
is \emph{no} transport at all along the backbone, and any walker is
confined forever to jump back-and-forth on the first long-range link
it accesses, which would imply $d_{w}=\infty$ in Eq.~(\ref{MSDeq}).
For any choice of $0<p<1$, no matter how close to unity $p$ gets,
at long-enough times the walker {}``escapes'' along the backbone
sufficiently often to explore ever-longer jumps both, to prevent confinement
and to exceed ordinary diffusion. Thus, we must conclude that even
this confinement fixed point is unstable and there has to be a third
fixed point, at least.

To find this fixed point, we have to move beyond looking at the stationary
behavior ($k=\infty$) of Eqs.~(\ref{eq:RG3RWfp}). Although Eqs.~(\ref{eq:RG3RWfp})
represent a five-dimensional parameter space, there do not appear
to be any further stationary points reachable from the initial conditions
in Eqs.~(\ref{eq:3parainit}) aside from those two already discussed.
As all flow appears to converge towards the singular confinement fixed
point, we make an Ansatz inserted into Eqs.~(\ref{eq:RG3RWfp}) for
$k\gg1$ that explores asymptotically the boundary layer \cite{BO}
in its vicinity:
\begin{eqnarray}
y^{(k)} & \sim & A_{y}\alpha^{-k}\quad\left(y\in\left\{
a,b,c,1-p_{1},1-p_{2}\right\} \right),
\label{eq:Ansatz0}
\end{eqnarray}
with the assumption that $\alpha>1$. Expanding Eqs.~(\ref{eq:RG3RWfp})
to leading order in $\alpha^{-k}$, we find an over-determined system
of equations

\begin{eqnarray}
\frac{1}{\alpha}A_{a} & = & \frac{1}{\alpha}A_{b}=\frac{A_{a}^{2}}{A_{1-p_{1}}}+A_{c},\nonumber \\
\frac{1}{\alpha}A_{c} & = & \frac{A_{a}^{2}}{2A_{1-p_{1}}},\label{eq:A}\\
\frac{1}{\alpha}A_{1-p_{1}} & = & A_{1-p_{2}}-\frac{2A_{a}^{2}}{A_{1-p_{1}}},\nonumber \\
\frac{1}{\alpha}A_{1-p_{2}} & = & A_{1-p_{1}}-\frac{A_{a}^{2}}{A_{1-p_{1}}}.\nonumber \end{eqnarray}
Exercising the freedom to choose $A_{a}=1$, we find\begin{eqnarray}
A_{b} & = & A_{a}=1,\nonumber \\
A_{c} & = & \frac{1}{2\phi},\nonumber \\
A_{1-p_{1}} & = & 2,\label{eq:As}\\
A_{1-p_{2}} & = & \phi^{2},\nonumber 
\end{eqnarray}
where 
\begin{equation}
\phi=\frac{\sqrt{5}+1}{2}=1.6180\ldots
\label{eq:goldensection}
\end{equation}
is the {}``golden ratio'' \cite{Livio03}, and obtain the eigenvalue equation
\begin{equation}
\alpha^{3}\left(\alpha+3\right)=8.
\label{eq:eigenvalue}
\end{equation}
It has a unique solution that satisfies the condition on the boundary
layer, $\alpha>1$, namely
\begin{eqnarray}
\alpha & = & \frac{2}{\phi}=1.2361\ldots.
\label{eq:alpha1}
\end{eqnarray}

Thus, we found another fixed point, lurking in the boundary layer
of the confinement fixed point but with very distinct physical properties
from it. Each reduction of the system size by a factor of 2 is accompanied
by a rescaling of the hopping parameters by a factor of $\alpha^{-1}$,
bringing them closer to confinement, yet, leaving just enough room
to escape and find still longer jumps. In fact, at this point of the
analysis it is not even obvious whether these ever less frequent escapes
from total confinement ultimately would result in sub-diffusive, normal,
or super-diffusive behavior.

\begin{figure}
\includegraphics[scale=0.32]{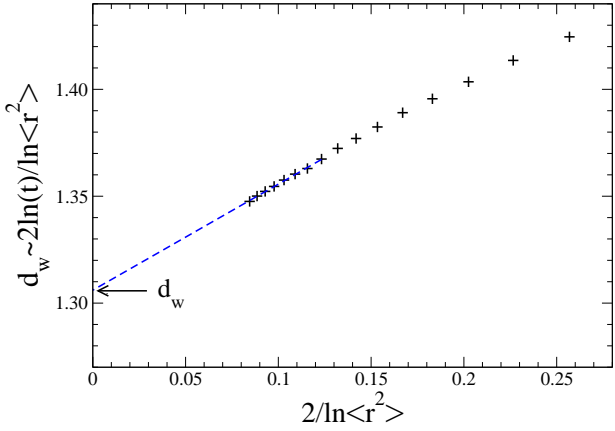} 
\caption{Plot of the results from simulations of the mean-square displacement
of random walks on HN3 displayed in Fig.~\ref{fig:3hanoi}. More
than $10^{7}$ walks were evolved up to $t_{{\rm max}}=10^{6}$ steps
to measure $\langle r^{2}\rangle_{t}$. The data is extrapolated according
to Eq.~(\ref{MSDeq}), such that the intercept on the vertical axis
determines $d_{w}$ asymptotically. The exact result from Eq.~(\ref{eq:D-expo})
is indicated by the arrow. }
\label{fig:MSDextra} 
\end{figure}

The leading-order Ansatz in Eq.~(\ref{eq:Ansatz0}) merely provided
the existence of a third fixed point at infinite times. Extending
the analysis to include finite-time corrections (i.~e., $\epsilon=1-z\ll1$),
we include first-order corrections, 
\begin{eqnarray}
y^{(k)} & \sim & A_{y}\alpha^{-k}\left\{ 1+\epsilon
B_{y}\beta^{k}+\ldots\right\} ,
\label{eq:Ansatz}
\end{eqnarray}
expand the recursions in (\ref{eq:RG3RWfp}) in $\epsilon$, and then
also use the fact that $\alpha^{-k}$ is small. Using the leading-order
constants $A_{y}$ in Eqs.~(\ref{eq:As}), also the next-leading constants
$B_{y}$ are determined self-consistently. We extract, again uniquely,
\begin{equation}
\beta=2\alpha=\frac{4}{\phi}
\label{eq:beta}
\end{equation}
and find, choosing $B_{a}=1$, 
\begin{eqnarray}
B_{b} & = & B_{a}=1,\nonumber \\
B_{c} & = & \frac{4}{5}\,\phi,\nonumber \\
B_{1-p_{1}} & = & -\frac{6}{5},\label{eq:Bs}\\
B_{1-p_{2}} & = & -\frac{8}{5\phi^{2}}.\nonumber 
\end{eqnarray}
Accordingly, time rescales as 
\begin{eqnarray}
T & \to & T'=\frac{4}{\phi}\, T,
\label{eq:Tscal}
\end{eqnarray}
and we obtain from Eq.~(\ref{MSDeq}) with $T\sim L^{d_{w}}$ for
the diffusion exponent for HN3: 
\begin{eqnarray}
d_{w} & = & 2-\log_{2}\phi=1.30576\ldots.
\label{eq:D-expo}
\end{eqnarray}
Our simulations in Fig.~\ref{fig:MSDextra} are in excellent agreement
with this result for $d_{w}$.

To extract the scaling behavior of the first-return distribution $Q_{0}(t)$
defined in Eq.~(\ref{eq:Qt}), we proceed similar to the discussion
in Sec. \ref{sub:RG-for-2RW}. Here, we will find that different sites
behave differently with respect to their return-time behavior, which
is not surprising as the hierarchy of long-range jumps restricts translational
invariance along the backbone. As in the discussion of first returns
in Sec. \ref{sub:RG-for-2RW}, we anticipate that we need an expansion
of the parameters to order $\epsilon^{2}$, i. e., we extend the Ansatz
in Eq.~(\ref{eq:Ansatz}) even further. We find that it is sufficient
to use
\begin{eqnarray}
y^{(k)} & \sim & A_{y}\alpha^{-k}\left\{ 1+\epsilon B_{y}\beta^{k}+\epsilon^{2}C_{y}\beta^{2k}+\ldots\right\}\nonumber \\
 & \sim & A_{y}\left\{ \alpha^{-k}+\epsilon
B_{y}2^{k}+\epsilon^{2}C_{y}\left(4\alpha\right)^{k}+\ldots\right\} 
\label{eq:Ansatz2}
\end{eqnarray}
with
\begin{eqnarray}
C_{a} & = & C_{b}=\frac{121}{25}\,\frac{\phi}{16},\nonumber \\
C_{c} & = & \frac{121}{25}\,\frac{\phi^{2}}{16},\nonumber \\
C_{1-p_{1}} & = & -\frac{121}{25}\,\frac{\phi}{16},\label{eq:Cs}\\
C_{1-p_{2}} & = & -\frac{121}{25}\,\frac{1}{16\phi},\nonumber 
\end{eqnarray}
where we have also dropped at each order in $\epsilon$ terms that
grow less than the leading exponential in $k$. For the coefficients
$C_{y}$ there is no freedom to choose, as they are a result of quadratic
terms of the previous order, $\epsilon B_{y}\beta^{k}$; other terms
with that freedom are subdominant.

As for Eq.~(\ref{eq:1d-elementary}), after $k=K-1$ RG steps, the
system has reduced to an elementary graph consisting of a single segment
like that one shown in Fig.~\ref{fig:RG3RW} (left) with $n=2$ but
with $p_{2}=0$. Now, all even sites ($l=0,2,4$) are no longer connected
to a long-range jump. (Jumps of rate $c$ do not count as long-range,
since $c\to0$ for $k\to\infty$, whereas $p_{1}\sim p_{2}\to1$.)
We therefore consider two (extreme) possibilities: (1) Returns to
a starting point at the boundary ($l=0,N$) or central site ($l=N/2$)
on the network, and (2) returns to a site ($l=N/4$ or $l=3N/4$)
with the longest-possible long-range jump on the original network. 

\begin{figure}
\includegraphics[scale=0.32]{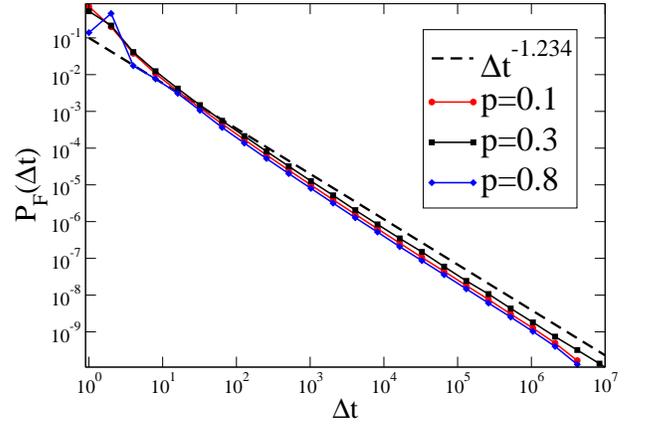} 
\caption{Plot of the probability $Q(\Delta t)$ of first returns to the origin
after $\Delta t$ update steps on a system of unlimited size. Data
was collected for three different walks on HN3 with $p=0.1$ (circles),
$p=0.3$ (squares), and $p=0.8$ (diamonds). The data with the smallest
and largest $p$ exhibit strong transient effects. The exact result
in Eq.~(\ref{eq:3RWmu}), $\mu=1.234\ldots$, is indicated by the
dashed line. }
\label{fig:FR} 
\end{figure}

First, we consider case (1) of starting at a boundary site, say, $l=0$.
It is easy to show that the central site $l=N/2$ behaves identical
to those on the boundary, and furthermore, even making the boundary
sites more accessible be using periodic boundary conditions does not
change the conclusion. We solve the system of equations 
\begin{eqnarray}
\tilde{P}_{0} & = & a\,\tilde{P}_{1}+c\,\tilde{P}_{2}+1,\nonumber \\
\tilde{P}_{1} & = & 2b\,\tilde{P}_{0}+b\,\tilde{P}_{2}+p_{1}\,\tilde{P}_{3},\nonumber \\
\tilde{P}_{2} & = & a\left[\tilde{P}_{1}+\tilde{P}_{3}\right]+2c\left[\tilde{P}_{0}+\tilde{P}_{4}\right],\label{eq:3RW-elemetary}\\
\tilde{P}_{3} & = & b\,\tilde{P}_{2}+2b\,\tilde{P}_{4}+p_{1}\,\tilde{P}_{1},\nonumber \\
\tilde{P}_{4} & = & a\,\tilde{P}_{3}+c\,\tilde{P}_{2},\nonumber 
\end{eqnarray}
where we have suppressed the superscript $^{(K-1)}$ on the generating
function and the parameters alike. {[}As we have learned from the
$1d$-walk in Eq.~(\ref{eq:1d-elementary}), the hopping parameters
originating from the boundary sites $l=0$ and $l=4$ have to be doubled
due to the reflecting boundaries.] The solution for $\tilde{P}_{0}$
of this system of equations yields
\begin{eqnarray}
&&\tilde{P}_{0} =\label{eq:3RW-P0} \\
&&\frac{\left(1-2c^{2}\right)\left(1-p_{1}^{2}\right)-2ab\left[1+\left(1+2c\right)\left(1+p_{1}\right)\right]+2a^{2}b^{2}}{\left[\left(1-p_{1}\right)\left(1-2c\right)-4ab\right](1+2c)\left(1+p_{1}-2ab\right)}.\nonumber
\end{eqnarray}
Notice that \emph{both,} numerator and denominator, decay asymptotically
like $\sim1-p_{1}^{2}$ near the stable fixed point. Dividing out
that behavior, both behave as $1+y^{(K)}$ with $y^{(K)}$ as in Eq.
(\ref{eq:Ansatz2}), where $y$ here stands for a mix of coefficient
that results from multiplying out the terms in numerator and denominator,
respectively. Since we do not expect any spurious cancellations, the
ratio of these two expressions \emph{also} results in the form $1+y^{(K)}$.
Hence, making superscripts reappear, we find from Eq.~(\ref{eq:Ansatz2})
\begin{eqnarray}
\frac{1}{\tilde{P}_{0}^{(K-1)}} & \sim & 1+{\cal
  B}\,\epsilon\,2^{K}+{\cal
  C}\,\epsilon^{2}\left(4\alpha\right)^{K}+\ldots,
\label{eq:3RW-Porigin}
\end{eqnarray}
where we have also dropped the $\alpha^{-K}$-term in order $\epsilon^{0}$
and marked constants that are unimportant for the scaling with $K$
by calligraphy script. Using Eq.~(\ref{eq:3RWampli-reno}) inserted
into the relation for $\tilde{Q}_{0}$ in Eq.~(\ref{eq:Q}), we find
\begin{eqnarray}
\tilde{Q}_{0} & = & 1-\frac{1}{\tilde{P}_{0}^{(0)}}\nonumber \\
 & = &
1-\frac{1}{\tilde{P}_{0}^{(K-1)}}\,\prod_{k=0}^{K-2}\left[1-\frac{2a^{(k)}b^{(k)}}{1-\left(p_{1}^{(k)}\right)^{2}}\right].
\label{eq:3RW-Qorigin}
\end{eqnarray}
Ignoring at most a finite number of factors in the product (hence,
missing an overall constant ${\cal A}$), we can expand the remaining
factors in the product using the asymptotic expansion in Eq.~(\ref{eq:Ansatz2})
for $1\ll k\leq K-2\to\infty$: 
\begin{eqnarray}
&&\prod_{k=0}^{K-2}\left[1-\frac{2a^{(k)}b^{(k)}}{1-\left(p_{1}^{(k)}\right)^{2}}\right] \nonumber \\
 && \sim  {\cal A}\prod_{k\gg1}^{K-2}\left[1-\alpha^{-k}+{\cal D}\,\epsilon\,2^{k}+{\cal E}\epsilon^{2}\left(4\alpha\right)^{k}+\ldots\right],\nonumber \\
 & &\sim  {\cal A}\left[1-{\cal D}\,\epsilon\sum_{k\gg1}^{K-2}2^{k}+{\cal E}\,\epsilon^{2}\sum_{k\gg1}^{K-2}\left(4\alpha\right)^{k}+\ldots\right],\nonumber \\
 & &\sim  {\cal A}+{\cal F}\,\epsilon\,2^{K}+{\cal
    G}\,\epsilon^{2}\left(4\alpha\right)^{K}+\ldots.
\label{eq:Product}
\end{eqnarray}
Note that $0<{\cal A}<1$, as each factor in the product must be between
$\frac{1}{2}$ and 1 for any choice the probability $p$. Inserting
Eqs.~(\ref{eq:3RW-Porigin}) and~(\ref{eq:Product}) into Eq.~(\ref{eq:3RW-Qorigin}),
we obtain
\begin{eqnarray}
\tilde{Q}_{0} & \sim & \left(1-{\cal A}\right)+{\cal
  H}\,\epsilon\,2^{K}+{\cal
  I}\,\epsilon^{2}\left(4\alpha\right)^{K}+\ldots.
\label{eq:non-recurQ}
\end{eqnarray}
This result implies first that returns to sites like $l=0,N/2,N$
are \emph{not} recurrent because walks may escape forever with a finite
probability ${\cal A}$. Finally, an application of Eq.~(\ref{eq:def-moments})
produces
\begin{eqnarray*}
\left\langle t\right\rangle _{K} & \sim & 2^{K}\sim N,\\
\left\langle t^{2}\right\rangle _{K} & \sim &
\left(4\alpha\right)^{K}\sim N^{\log_{2}\left(4\alpha\right)},
\end{eqnarray*}
hence, with Eq.~(\ref{eq:mu}) and $\alpha=2/\phi$:
\begin{eqnarray}
\mu & = & 2-\frac{1}{2-\log_{2}\left(\phi\right)}=1.23416\ldots.
\label{eq:3RWmu}
\end{eqnarray}
A relation of this form, $\mu=2-1/d_{w}$, is commonly found for L\'evy
flights~\cite{Metzler04}, and the result is again borne out by
our simulations, see Fig.~\ref{fig:FR}.

Case (2), referring to walkers starting near the longest jump in the
network, is represented by the system 
\begin{eqnarray}
\tilde{P}_{0} & = & a\,\tilde{P}_{1}+c\,\tilde{P}_{2},\nonumber \\
\tilde{P}_{1} & = & 2b\,\tilde{P}_{0}+b\,\tilde{P}_{2}+p_{1}\,\tilde{P}_{3}+1,\nonumber \\
\tilde{P}_{2} & = & a\left[\tilde{P}_{1}+\tilde{P}_{3}\right]+2c\left[\tilde{P}_{0}+\tilde{P}_{4}\right],\label{eq:3RW-elemetary-long}\\
\tilde{P}_{3} & = & b\,\tilde{P}_{2}+2b\,\tilde{P}_{4}+p_{1}\,\tilde{P}_{1},\nonumber \\
\tilde{P}_{4} & = & a\,\tilde{P}_{3}+c\,\tilde{P}_{2},\nonumber 
\end{eqnarray}
with the constant term from the initial conditions now near an odd
site, say, $l=1$. Solving for the launch site of the walk, we find
\begin{eqnarray}
\tilde{P}_{1} & = &
\frac{1-2c-ab\left(3-2c\right)}{\left[\left(1-p_{1}\right)\left(1-2c\right)-4ab\right]\left(1+p_{1}-2ab\right)}.
\label{eq:3RW-P1}
\end{eqnarray}
Note that in this case \emph{only} the denominator of
$\tilde{P}_{1}=\tilde{P}_{1}^{(K-1)}$ decays with $1-p_{1}$ on
approaching the stable fixed point which will ensure that
$\tilde{P}_{N/4}^{(0)}$ diverges, making this process recurrent on
behalf of Eq.~(\ref{eq:Q}). In contrast to Eq.~(\ref{eq:3RW-Porigin}),
we find here
\begin{eqnarray}
&&\frac{1}{\tilde{P}_{1}^{(K-1)}}
\label{eq:3RW-P-long}\\
 && \sim  \left(1-p_{1}\right)\left[1+{\cal J}\,\epsilon\,2^{K}+{\cal K}\epsilon^{2}\left(4\alpha\right)^{K}+\ldots\right],\nonumber\\
 && \sim  A_{1-p_{1}}\alpha^{-K+1}\left[1+{\cal J}\,\epsilon\,2^{K}+{\cal K}\epsilon^{2}\left(4\alpha\right)^{K}+\ldots\right]\nonumber\\
 && \quad\left[1+\epsilon B_{1-p_{1}}\left(2\alpha\right)^{K-1}+\epsilon^{2}C_{1-p_{1}}\left(2\alpha\right)^{2K-2}+\ldots\right],\nonumber\\
 && \sim  \left(2\alpha\right)\alpha^{-K}\left[1+{\cal
      L}\,\epsilon\,\left(2\alpha\right)^{K}+{\cal
      M}\,\epsilon^{2}\left(2\alpha\right)^{2K}+\ldots\right],\nonumber
\end{eqnarray}
where the last step is justified by the observation that the leading
terms order-by-order in $\epsilon$ in the first square bracket dominate
over those in the second. 

We have to apply Eq.~(\ref{eq:3RWampli-reno2}) for this case to obtain
\begin{eqnarray*}
\tilde{P}_{\frac{N}{4}}^{(0)} & = & \tilde{P}_{1}^{(K-1)}\,\prod_{k=0}^{K-2}\frac{1-p_{1}^{(k)}}{1-p_{1}^{(k)}-2a^{(k)}b^{(k)}}\end{eqnarray*}
and from Eq.~(\ref{eq:Q}):\begin{eqnarray}
\tilde{Q}_{\frac{N}{4}} & = & 1-\frac{1}{\tilde{P}_{\frac{N}{4}}^{(0)}},\nonumber \\
 & = & 1-\frac{1}{\tilde{P}_{1}^{(K-1)}}\,\prod_{k=0}^{K-2}\left[1-\frac{2a^{(k)}b^{(k)}}{1-p_{1}^{(k)}}\right].\label{eq:QN4}\end{eqnarray}
The product in Eq.~(\ref{eq:QN4}) behaves identically to that discussed
above in Eq.~(\ref{eq:Product}) and hence will not alter the overall
form of the expansion when multiplying $1/\tilde{P}_{1}^{(K-1)}$
in Eq.~(\ref{eq:3RW-P-long}). Then we get\begin{eqnarray}
\tilde{Q}_{\frac{N}{4}} & \sim & 1-\left(2\alpha\right)\alpha^{-K}\left[1+{\cal N}\,\epsilon\,\left(2\alpha\right)^{K}+{\cal O}\epsilon^{2}\left(2\alpha\right)^{2K}+\ldots\right],\nonumber \\
 & \sim & 1-\left(2\alpha\right)\alpha^{-K}+{\cal P}\,\epsilon\,2^{K}+{\cal Q}\,\epsilon^{2}\left(4\alpha\right)^{K}+\ldots.\label{eq:Q-recur}\end{eqnarray}
This result is almost identical to that for returns to the origin
above in Eq.~(\ref{eq:non-recurQ}), except that sites at the longest
jump in the system are recurrent, i. e., $\tilde{Q}_{N/4}\equiv1$
for large systems ($K\to\infty$) and times ($\epsilon=1-z\to0$).
(Presumably recurrence will gradually degrade from the strictly recurrent
sites at the highest level in the hierarchy to those at the lowest.)
Yet, the scaling of return times in form of the exponent $\mu$ is
described by Eq.~(\ref{eq:3RWmu}) for \emph{all} sites.

\subsection{Generating Function for Random Walks on HN4\label{sub:RG-for-4RW}}

Next, we consider a random walk on HN4. The {}``master-equation''~\cite{Redner01}
for the probability of the walker to be at site $n$, as defined in
Eq.~(\ref{eq:numbering}), at time $t$ is given by 
\begin{eqnarray}
P_{n,t} & = & \frac{1-p}{2}\left[P_{n-1,t-1}+P_{n+1,t-1}\right]\nonumber \\ &  &
\quad+\frac{p}{2}\left[P_{n-2^{i+1},t-1}+P_{n+2^{i+1},t-1}\right],
\label{eq:4RW}
\end{eqnarray}
where $p$ is the probability to make a long-range jump. (Throughout,
we considered $p$ uniform, independent of $n$ or $t$). To make the
connection between $n$ and $i$ explicit, we rewrite Eq.~(\ref{eq:4RW})
as
\begin{eqnarray}
P_{2^{i}\left(2j+1\right),t} & = &
\frac{1-p}{2}\left[P_{2^{i}\left(2j+1\right)-1,t-1}+P_{2^{i}\left(2j+1\right)+1,t-1}\right]\nonumber\\ 
&  &+\frac{p}{2}\left[P_{2^{i}\left(2j-1\right),t-1}+P_{2^{i}\left(2j+3\right),t-1}\right],\nonumber \\
P_{0,t} & = & \frac{1-p}{2}\left[P_{-1,t-1}+P_{1,t-1}\right]+p\,
P_{0,t-1}.
\label{eq:new4RW}
\end{eqnarray}
Note that the case $n=0$ is not covered by Eq.~(\ref{eq:numbering})
and, hence, must be treated separately. Here, we choose the site $n=0$
to be the only one connected to itself such that HN4 is 4-regular
throughout, as depicted in Fig.~\ref{fig:4hanoi}.

It is straightforward to apply the generating function in
Eq.~(\ref{eq:generator}) again, assuming, for simplicity, the initial
condition
\begin{eqnarray}
P_{n,0} & = & \delta_{n,0}.
\label{eq:RW4init}
\end{eqnarray}
We obtain
\begin{eqnarray}
\tilde{P}{}_{2^{i}\left(2j+1\right)} & = & \frac{1-p}{2}\, z\,\left[\tilde{P}_{2^{i}\left(2j+1\right)-1}+\tilde{P}_{2^{i}\left(2j+1\right)+1}\right]\nonumber\\ 
&  &+\frac{p}{2}\, z\,\left[\tilde{P}_{2^{i}\left(2j-1\right)}+\tilde{P}_{2^{i}\left(2j+3\right)}\right],\nonumber \\
\tilde{P}_{0}-1 & = & \frac{1-p}{2}\,
z\,\left[\tilde{P}_{-1}+\tilde{P}_{1,}\right]+p\, z\,\tilde{P}_{0}.
\label{eq:new4RWgen}
\end{eqnarray}
While the overall structure of this problem is even more symmetric
than for HN3 in Sec. \ref{sub:RG-for-3RW}, a RG treatment does not
seem possible in this case. Tracing out all odd sites would immediately
interconnect \emph{all} other remaining sites. (The resulting infinite
set of coupled equations may have certain symmetry properties that
would lend themselves for a recursive treatment. We have not yet explored
such a possibility.) 

In contrast to an ordinary lattice, say, it is also not straightforward
to solve this equation by a Fourier transform such as 
\begin{eqnarray}
F(z,\phi) & = & \sum_{n=-\infty}^{\infty}\tilde{P}_{n}(x)\, e^{n\phi
  I},
\label{eq:Fourier}
\end{eqnarray}
defining $I=\sqrt{-1}$. Considering the $1-p$ terms, originating
from nearest-neighbor jumps, and the $p$ terms, originating from
long-range jumps, in Eqs.~(\ref{eq:new4RWgen}) separately provides
for regularly-space patterns, level-by-level in the hierarchy. But
the mixing of nearest-neighbor and long-range jumps destroys this
regularity. Hence, we resort to transforming the Eqs.~(\ref{eq:new4RWgen})
in each level $i$ with a partial transform,
\begin{equation}
\Pi_{i}(z,\phi) = 
\sum_{j=-\infty}^{\infty}\tilde{P}_{2^{i}(2j+1)}(z)\,\exp\left\{
2^{i}\left(2j+1\right)\phi I\right\} .
\label{eq:level-transform}
\end{equation}
Inserting Eq.~(\ref{eq:level-transform}) and application of the general
theorem
\begin{eqnarray}
\sum_{i=1}^{\infty}\sum_{j=-\infty}^{\infty}f_{2^{i}\left(2j+1\right)\pm1}
& = & \sum_{j=-\infty}^{\infty}f_{2j+1}-f_{\pm1},
\label{eq:theorem}
\end{eqnarray}
which results because $2^{i}(2j+1)$ for $i\geq1$ exactly runs over
all even numbers $\not=0$ and over all odd numbers for $i=0$, yields
for Eqs.~(\ref{eq:new4RWgen}):
\begin{eqnarray}
&&\sum_{i=1}^{\infty}\Pi_{i}\left[1-pz\cos\left(2^{i+1}\phi\right)\right]\nonumber\\ 
&& \quad =  (1-p)z\cos\phi\Pi_{0}+1-(1-pz)\tilde{P}_{0},\nonumber \\
\nonumber \\
&&\Pi_{0}\left[1-pz\cos\left(2\phi\right)\right]
\label{eq:4RWtransformed}\\ 
&& \quad =
(1-p)z\cos\phi\sum_{i=1}^{\infty}\Pi_{i}+(1-p)z\cos\phi\tilde{P}_{0}.\nonumber
\end{eqnarray}
We can combine both relations to get
\begin{eqnarray}
1&=&\left[1-zp-(1-p)z\cos\phi\right]F\nonumber\\ 
&& \quad+zp\sum_{i=0}^{\infty}\left[1-\cos\left(2^{i+1}\phi\right)\right]\Pi_{i},
\label{eq:F-Pi}
\end{eqnarray}
using Eqs.~(\ref{eq:Fourier}) and (\ref{eq:level-transform}) to
eliminate $\tilde{P}_{0}$ via
\begin{eqnarray}
\sum_{i=0}^{\infty}\Pi_{i} & = & \sum_{i=0}^{\infty}\sum_{j=-\infty}^{\infty}\tilde{P}_{2^{i}(2j+1)}\,\exp\left\{ 2^{i}\left(2j+1\right)\phi I\right\} ,\nonumber\\
 & = & \sum_{n=-\infty}^{\infty}\tilde{P}_{n}\, e^{n\phi I}-\tilde{P}_{0},\\
 & = & F-\tilde{P}_{0}.\nonumber
\end{eqnarray}
At this point, there does not seem to be any further progress possible
on Eq.~(\ref{eq:F-Pi}), due to the term $\sum_{i}\Pi_{i}\cos\left(2^{i+1}\phi\right)$,
which resembles a Weierstrass function \cite{Hughes81}. At best,
on could try to extract information about the moments of the walk,
\begin{eqnarray*}
\left\langle n^{k}\right\rangle _{t} & = & \sum_{n=-\infty}^{\infty}n^{k}P_{n,t},\end{eqnarray*}
via the moment-generating function\begin{eqnarray}
M_{k}(z) & = & \sum_{t=0}^{\infty}\left\langle n^{k}\right\rangle _{t}\, z^{t},\nonumber \\
 & = & \left[-I\partial_{\phi}\right]^{k}F(z,\phi)|_{\phi=0}.\label{eq:Moment-gen}\end{eqnarray}
Note that the 2nd moment $M_{2}(z)$ already would provide the exponent
$d_{w}$ on behalf of the definition in Eq.~(\ref{MSDeq}). {[}All
odd moments vanish, of course, as Eq.~(\ref{eq:F-Pi}) is even in
$\phi$.] The 0th moment, setting $\phi=0$ in Eq.~(\ref{eq:F-Pi}),
simply results in \begin{eqnarray*}
M_{0}(z) & = & \frac{1}{1-z},\end{eqnarray*}
which just demonstrates that everything is properly normalized, $ $${\cal N}_{t}=\left\langle n^{0}\right\rangle _{t}=\sum_{n=-\infty}^{\infty}P_{n,t}=1$,
at all times $t$. But already the 2nd moment would lead to terms
containing $\sum_{i}\Pi_{i}4^{i}$, which we can not account for,
even at $\phi=0$ and in the limit $z\to1^{-}$.

Instead, we note that the long-time behavior is dominated by the long-range
jumps, as discussed for HN3 in Sec. \ref{sub:RG-for-3RW}. To simplify
matters, we set $p=1/2$ here, although any finite probability would
lead to the same conclusions. We make an {}``annealed'' approximation,
i.~e., we assume that we happen to be at some site $n$ in Eq.~(\ref{eq:numbering})
with probability $1/2^{i}$, corresponding to the relative frequency
of such a site, yet independent of time or history. This ignores the
fact that in the network geometry a long jump of length $2^{i}$ can
be followed \emph{only} by another jump of that length or a jump of
unit length, and that many intervening steps are necessary to make
a jump of length $2^{i+1}$, for instance. Here, at each instant the
walker jumps a distance $2^{i}$ left or right irrespectively with
probability $1/2^{i+1}$, and we can write \begin{eqnarray}
{\cal P}_{n,t} & = & \sum_{n'}T_{n,n'}{\cal P}_{n',t-1}\label{eq:Transfer}\end{eqnarray}
 with \begin{eqnarray}
T_{n,n'} & = & \frac{a-1}{2a}\sum_{i=0}^{\infty}a^{-i}\left(\delta_{n-n',b^{i}}+\delta_{n-n',-b^{i}}\right),\label{eq:T}\end{eqnarray}
 where $a=b=2$. Eqs.~(\ref{eq:Transfer}-\ref{eq:T}) are identical
to the Weierstrass random walk discussed in Refs.~\cite{Hughes81,Shlesinger93}
for arbitrary $1<a<b^{2}$. There, it was shown that $d_{w}=\ln(a)/\ln(b)$,
which leads to the conclusion that $d_{w}=1$ in Eq.~(\ref{MSDeq})
for HN4, as has been predicted (with logarithmic corrections) on the
basis of numerical simulations in Ref.~\cite{SWPRL}.

\section{Conclusions\label{sec:Conclusions}}

We have show how the powerful tools of the dynamic renormalization
group \cite{Redner01} allow to dissect this intricate random walk
problem on the planar network HN3 with a {}``hidden'' fixed point.
Indeed, using a boundary-layer analysis, we unravel the irregular
singularity of the dominant fixed point in a five-dimensional parameter
space, resulting in a set of exact, non-trivial exponents describing
super-diffusive transport. Adding just one more link to each site,
we obtain a non-planar network HN4 which possess an even higher degree
of symmetry, yet, for which we can only develop an equation for the
generator and an alternative {}``annealed'' treatment which provides
results that are consistent with simulations. (We believe that a proper
exploitation of the symmetry in HN4, which eludes us here, will ultimately
make exact results possible.) 

Aside from the singular fixed point, HN3 serves further as an instructive
example for a network in which nodes have heterogeneous recurrence
properties. The diffusion exponent $d_{w}$ is larger than the fractal
dimension $d_{f}=1$ of the lattice backbone that the walk is embedded
in, which usually implies recurrence \cite{Bollt05,Condamin07}.
Here, the near-confined state of the walk favors recurrences to sites
in higher levels of the hierarchy, although the associated first-return
exponent is the same for all sites for the time distribution of any
given return.

We should also mention that our results for HN3 can have an alternative
interpretation. If we ignore the one-dimensional lattice backbone
and instead consider the network as graph without particular embedding,
then Eq.~(\ref{eq:3dia}) for the diameter, or more specifically the
average growth in neighborhood $S_{d}\sim d^{2}$ with jump-distance
$d$ found in Fig.~\ref{fig:HN3neighbors}, implies that the fractal
dimension for that graph is $d_{f}=2$. The RG would discover the
then-obscured asymmetry between the backbone and long-range jumps
(even when starting with $p=1/3$) and lead to the same analysis.
Yet, with all distances now being (on average) measured as the square-root
of their separation along the backbone, also the mean-square displacement
in Eq.~(\ref{MSDeq}) needs to be reevaluated, yielding a diffusion
exponent twice its previous value, $d_{w}=2(2-\log_{2}\phi)=2.61\ldots$.
In this interpretation, $d_{w}>d_{f}$ still applies, but walks are
now sub-diffusive in this measure. Of course, an exponent that is
independent of such a metric, like the purely event-base first return
probability, does \emph{not} change. In turn, the relation between
$d_{w}$ and $\mu$ fails, consistent with the fact that the walk
can no longer be considered a L\'evy flight.

Finally, our results suggest that many other interesting transport
phenomena, such as voter models, exclusion processes, or self-organized
critical phenomena can be fruitfully studied on these networks, which
are sufficiently complex for interesting results but sufficiently
simple to be tractable. Especially in light of the tremendous interest
in complex dynamics on designed structures, we hope that these networks
can make a useful contribution~\cite{Barabasi01,Andrade05,Hinczewski06,Zhang07}.

\bibliographystyle{apsrev}
\bibliography{/Users/stb/Boettcher}

\begin{thebibliography}{22}
\expandafter\ifx\csname natexlab\endcsname\relax\def\natexlab#1{#1}\fi
\expandafter\ifx\csname bibnamefont\endcsname\relax
  \def\bibnamefont#1{#1}\fi
\expandafter\ifx\csname bibfnamefont\endcsname\relax
  \def\bibfnamefont#1{#1}\fi
\expandafter\ifx\csname citenamefont\endcsname\relax
  \def\citenamefont#1{#1}\fi
\expandafter\ifx\csname url\endcsname\relax
  \def\url#1{\texttt{#1}}\fi
\expandafter\ifx\csname urlprefix\endcsname\relax\def\urlprefix{URL }\fi
\providecommand{\bibinfo}[2]{#2}
\providecommand{\eprint}[2][]{\url{#2}}

\bibitem[{\citenamefont{{Erd\"os} and {R\'enyi}}(1973)}]{ER}
\bibinfo{author}{\bibfnamefont{P.}~\bibnamefont{{Erd\"os}}} \bibnamefont{and}
  \bibinfo{author}{\bibfnamefont{A.}~\bibnamefont{{R\'enyi}}}, in
  \emph{\bibinfo{booktitle}{The Art of Counting}} (\bibinfo{address}{MIT,
  Cambridge}, \bibinfo{year}{1973}).

\bibitem[{\citenamefont{Bollobas}(1985)}]{Bollobas}
\bibinfo{author}{\bibfnamefont{B.}~\bibnamefont{Bollobas}},
  \emph{\bibinfo{title}{Random Graphs}} (\bibinfo{publisher}{Academic Press},
  \bibinfo{address}{London}, \bibinfo{year}{1985}).

\bibitem[{\citenamefont{Berker and Ostlund}(1979)}]{Berker79}
\bibinfo{author}{\bibfnamefont{A.~N.} \bibnamefont{Berker}} \bibnamefont{and}
  \bibinfo{author}{\bibfnamefont{S.}~\bibnamefont{Ostlund}},
  \bibinfo{journal}{Journal of Physics C: Solid State Physics}
  \textbf{\bibinfo{volume}{12}}, \bibinfo{pages}{4961} (\bibinfo{year}{1979}).

\bibitem[{\citenamefont{Migdal}(1976)}]{Migdal76}
\bibinfo{author}{\bibfnamefont{A.~A.} \bibnamefont{Migdal}},
  \bibinfo{journal}{J.~Exp.~Theo.~Phys.} \textbf{\bibinfo{volume}{42}},
  \bibinfo{pages}{743} (\bibinfo{year}{1976}).

\bibitem[{\citenamefont{Kadanoff}(1976)}]{Kadanoff76}
\bibinfo{author}{\bibfnamefont{L.~P.} \bibnamefont{Kadanoff}},
  \bibinfo{journal}{Ann.~Phys.} \textbf{\bibinfo{volume}{100}},
  \bibinfo{pages}{359} (\bibinfo{year}{1976}).

\bibitem[{\citenamefont{Mandelbrot}(1982)}]{Mandelbrot82}
\bibinfo{author}{\bibfnamefont{B.~B.} \bibnamefont{Mandelbrot}},
  \emph{\bibinfo{title}{The Fractal Geometry of Nature}}
  (\bibinfo{publisher}{Freeman, San Francisco}, \bibinfo{year}{1982}).

\bibitem[{\citenamefont{Barabasi et~al.}(2001)\citenamefont{Barabasi, Ravasz,
  and Vicsek}}]{Barabasi01}
\bibinfo{author}{\bibfnamefont{A.-L.} \bibnamefont{Barabasi}},
  \bibinfo{author}{\bibfnamefont{E.}~\bibnamefont{Ravasz}}, \bibnamefont{and}
  \bibinfo{author}{\bibfnamefont{T.}~\bibnamefont{Vicsek}},
  \bibinfo{journal}{Physica A} \textbf{\bibinfo{volume}{299}},
  \bibinfo{pages}{559} (\bibinfo{year}{2001}).

\bibitem[{\citenamefont{Andrade et~al.}(2005)\citenamefont{Andrade, Herrmann,
  Andrade, and da~Silva}}]{Andrade05}
\bibinfo{author}{\bibfnamefont{J.~S.} \bibnamefont{Andrade}},
  \bibinfo{author}{\bibfnamefont{H.-J.} \bibnamefont{Herrmann}},
  \bibinfo{author}{\bibfnamefont{R.~F.~S.} \bibnamefont{Andrade}},
  \bibnamefont{and} \bibinfo{author}{\bibfnamefont{L.~R.}
  \bibnamefont{da~Silva}}, \bibinfo{journal}{Phys. Rev. Lett.}
  \textbf{\bibinfo{volume}{94}}, \bibinfo{pages}{018702}
  (\bibinfo{year}{2005}).

\bibitem[{\citenamefont{Zhang et~al.}(2007)\citenamefont{Zhang, Zhou, Fang,
  Guan, and Zhang}}]{Zhang07}
\bibinfo{author}{\bibfnamefont{Z.}~\bibnamefont{Zhang}},
  \bibinfo{author}{\bibfnamefont{S.}~\bibnamefont{Zhou}},
  \bibinfo{author}{\bibfnamefont{L.}~\bibnamefont{Fang}},
  \bibinfo{author}{\bibfnamefont{J.}~\bibnamefont{Guan}}, \bibnamefont{and}
  \bibinfo{author}{\bibfnamefont{Y.}~\bibnamefont{Zhang}},
  \bibinfo{journal}{Europhysics Letters} \textbf{\bibinfo{volume}{79}},
  \bibinfo{pages}{38007} (\bibinfo{year}{2007}).

\bibitem[{\citenamefont{Southern and Young}(1977)}]{Southern77}
\bibinfo{author}{\bibfnamefont{B.~W.} \bibnamefont{Southern}} \bibnamefont{and}
  \bibinfo{author}{\bibfnamefont{A.~P.} \bibnamefont{Young}},
  \bibinfo{journal}{J.~Phys.~C: Solid State Phys.}
  \textbf{\bibinfo{volume}{10}}, \bibinfo{pages}{2179} (\bibinfo{year}{1977}).

\bibitem[{\citenamefont{Boettcher
  et~al.}(arxiv:0712.1259)\citenamefont{Boettcher, Gon{\c c}alves, and
  Guclu}}]{SWPRL}
\bibinfo{author}{\bibfnamefont{S.}~\bibnamefont{Boettcher}},
  \bibinfo{author}{\bibfnamefont{B.}~\bibnamefont{Gon{\c c}alves}},
  \bibnamefont{and} \bibinfo{author}{\bibfnamefont{H.}~\bibnamefont{Guclu}},
  \bibinfo{journal}{J. Phys. A: Math. Theo.} \textbf{\bibinfo{volume}{41}},
  \bibinfo{pages}{252001} (\bibinfo{year}{2008}).

\bibitem[{\citenamefont{Boettcher and Gon{\c c}alves}(arXiv:0802.2757)}]{SWN}
\bibinfo{author}{\bibfnamefont{S.}~\bibnamefont{Boettcher}} \bibnamefont{and}
  \bibinfo{author}{\bibfnamefont{B.}~\bibnamefont{Gon{\c c}alves}}
  (\bibinfo{year}{arxiv:0802.2757}).

\bibitem[{\citenamefont{Watts and Strogatz}(1998)}]{Watts98}
\bibinfo{author}{\bibfnamefont{D.~J.} \bibnamefont{Watts}} \bibnamefont{and}
  \bibinfo{author}{\bibfnamefont{S.~H.} \bibnamefont{Strogatz}},
  \bibinfo{journal}{Nature} \textbf{\bibinfo{volume}{393}},
  \bibinfo{pages}{440} (\bibinfo{year}{1998}).

\bibitem[{\citenamefont{Redner}(2001)}]{Redner01}
\bibinfo{author}{\bibfnamefont{S.}~\bibnamefont{Redner}},
  \emph{\bibinfo{title}{A Guide to First-Passage Processes}}
  (\bibinfo{publisher}{Cambridge University Press},
  \bibinfo{address}{Cambridge}, \bibinfo{year}{2001}).

\bibitem[{\citenamefont{Bender and Orszag}(1978)}]{BO}
\bibinfo{author}{\bibfnamefont{C.~M.} \bibnamefont{Bender}} \bibnamefont{and}
  \bibinfo{author}{\bibfnamefont{S.~A.} \bibnamefont{Orszag}},
  \emph{\bibinfo{title}{Advanced Mathematical Methods for Scientists and
  Engineers}} (\bibinfo{publisher}{McGraw-Hill}, \bibinfo{address}{New York},
  \bibinfo{year}{1978}).

\bibitem[{\citenamefont{Livio}(2003)}]{Livio03}
\bibinfo{author}{\bibfnamefont{M.}~\bibnamefont{Livio}},
  \emph{\bibinfo{title}{The Golden Ratio: The Story of PHI, the World's Most
  Astonishing Number}} (\bibinfo{publisher}{Broadway Books},
  \bibinfo{address}{New York}, \bibinfo{year}{2003}).

\bibitem[{\citenamefont{Metzler and Klafter}(2004)}]{Metzler04}
\bibinfo{author}{\bibfnamefont{R.}~\bibnamefont{Metzler}} \bibnamefont{and}
  \bibinfo{author}{\bibfnamefont{J.}~\bibnamefont{Klafter}},
  \bibinfo{journal}{J. Phys. A: Math. Gen.} \textbf{\bibinfo{volume}{37}},
  \bibinfo{pages}{R161} (\bibinfo{year}{2004}).

\bibitem[{\citenamefont{Hughes et~al.}(1981)\citenamefont{Hughes, Shlesinger,
  and Montroll}}]{Hughes81}
\bibinfo{author}{\bibfnamefont{B.~D.} \bibnamefont{Hughes}},
  \bibinfo{author}{\bibfnamefont{M.~F.} \bibnamefont{Shlesinger}},
  \bibnamefont{and} \bibinfo{author}{\bibfnamefont{E.~W.}
  \bibnamefont{Montroll}}, \bibinfo{journal}{Proc. Natl. Acad. Sci.}
  \textbf{\bibinfo{volume}{78}}, \bibinfo{pages}{3287} (\bibinfo{year}{1981}).

\bibitem[{\citenamefont{Shlesinger et~al.}(1993)\citenamefont{Shlesinger,
  Zaslavsky, and Klafter}}]{Shlesinger93}
\bibinfo{author}{\bibfnamefont{M.~F.} \bibnamefont{Shlesinger}},
  \bibinfo{author}{\bibfnamefont{G.~M.} \bibnamefont{Zaslavsky}},
  \bibnamefont{and} \bibinfo{author}{\bibfnamefont{J.}~\bibnamefont{Klafter}},
  \bibinfo{journal}{Natur} \textbf{\bibinfo{volume}{363}}, \bibinfo{pages}{31}
  (\bibinfo{year}{1993}).

\bibitem[{\citenamefont{Bollt and ben Avraham}(2005)}]{Bollt05}
\bibinfo{author}{\bibfnamefont{E.~M.} \bibnamefont{Bollt}} \bibnamefont{and}
  \bibinfo{author}{\bibfnamefont{D.}~\bibnamefont{ben Avraham}},
  \bibinfo{journal}{New Journal of Physics} \textbf{\bibinfo{volume}{7}},
  \bibinfo{pages}{26} (\bibinfo{year}{2005}).

\bibitem[{\citenamefont{Condamin et~al.}(2007)\citenamefont{Condamin, Benichou,
  Tejedor, Voituriez, and Klafter}}]{Condamin07}
\bibinfo{author}{\bibfnamefont{S.}~\bibnamefont{Condamin}},
  \bibinfo{author}{\bibfnamefont{O.}~\bibnamefont{Benichou}},
  \bibinfo{author}{\bibfnamefont{V.}~\bibnamefont{Tejedor}},
  \bibinfo{author}{\bibfnamefont{R.}~\bibnamefont{Voituriez}},
  \bibnamefont{and} \bibinfo{author}{\bibfnamefont{J.}~\bibnamefont{Klafter}},
  \bibinfo{journal}{Nature} \textbf{\bibinfo{volume}{450}}, \bibinfo{pages}{77}
  (\bibinfo{year}{2007}).

\bibitem[{\citenamefont{Hinczewski and Berker}(2006)}]{Hinczewski06}
\bibinfo{author}{\bibfnamefont{M.}~\bibnamefont{Hinczewski}} \bibnamefont{and}
  \bibinfo{author}{\bibfnamefont{A.}~\bibnamefont{Berker}},
  \bibinfo{journal}{Phys. Rev. E} \textbf{\bibinfo{volume}{73}},
  \bibinfo{pages}{066126} (\bibinfo{year}{2006}).

\end{thebibliography}

\end{document}